\documentclass{llncs}
\usepackage{braket}
\usepackage{amsmath}
\usepackage{amsfonts}
\usepackage{amssymb}

\usepackage{rotating}
\usepackage{mathtools}
\usepackage[bookmarks,bookmarksopen,bookmarksdepth=2]{hyperref}

\usepackage{algorithm}
\usepackage{algpseudocode}
\usepackage[utf8]{inputenc}
\usepackage{tikz}
\usepackage{pgfplots}
\usepackage{bbm}
\usepackage{multirow}
\usepackage{enumerate}
\usepackage{enumitem}
\usepackage{booktabs}
\usepackage{xcolor}
\usepackage{xifthen}
\usepackage{tikz}
\usepackage{subcaption}

\newif\ifeprint
\eprinttrue

\title{Quantum Period Finding against Symmetric Primitives in Practice}
\ifeprint
\author{Xavier Bonnetain\inst{1} \and Samuel Jaques\inst{2}}
\institute{Institute for Quantum Computing, Department of Combinatorics and Optimization, University of Waterloo, Canada\\
\email{xbonnetain@uwaterloo.ca}
\and
Department of Materials, University of Oxford, UK
\email{samuel.jaques@materials.ox.ac.uk}
}
\else
\author{}
\institute{}
\email{}
\fi

\newcommand{\zo}{\{0,1\}}
\newcommand{\measure}{\mathrlap{\frown}{\,\text{\scriptsize\rotatebox{20}{$\nearrow$}}}}
\newcommand{\midsection}{\tikz \draw[-] (-0.1,-0.1) -- (0.1,0.1);}

\usetikzlibrary{decorations.pathreplacing,arrows,patterns,calc,positioning,shapes.geometric,shapes.misc,intersections}
\tikzstyle{block}=[draw,minimum size=2em]
\tikzstyle{xor}=[draw, circle, thin,minimum size=0.8em,append after command={[shorten >=\pgflinewidth, shorten <=\pgflinewidth,inner sep=0] 
        (\tikzlastnode.north) edge (\tikzlastnode.south)
        (\tikzlastnode.east) edge (\tikzlastnode.west)
}]
\tikzstyle{l}=[minimum size=1.5em]
\tikzstyle{sponge}=[rectangle, rounded corners=.25cm, minimum width=.5cm, minimum height=1.8cm, draw]

\newcommand{\ceiling}[1]{\left\lceil#1\right\rceil}
\newfloat{circuit}{}{cir}
\floatname{circuit}{Circuit}

\algblockdefx{AAmp}{EndAA}[1]{\textbf{amplify} #1 \textbf{with}}{\textbf{end amplify}}

\newcommand{\Elephant}{\textsf{Elephant}}
\newcommand{\spongent}{\textsc{spongent}}
\newcommand{\keccak}{\textsc{Keccak}}
\newcommand{\Prince}{\textsf{PRINCE}}
\newcommand{\Chaskey}{\textsf{Chaskey}}

\begin{document}
\maketitle

\begin{abstract}
 We present the first complete implementation of the offline Simon's algorithm, and estimate its cost to attack the MAC Chaskey, the block cipher PRINCE and the NIST lightweight candidate AEAD scheme Elephant.
 
 These attacks require a reasonable amount of qubits, comparable to the number of qubits required to break RSA-2048. They are faster than other collision algorithms, and the attacks against PRINCE and Chaskey are the most efficient known to date. As Elephant has a key smaller than its state size, the algorithm is less efficient and ends up more expensive than exhaustive search.

We also propose an optimized quantum circuit for boolean linear algebra as well as complete reversible implementations of PRINCE, Chaskey, spongent and Keccak which are of independent interest for quantum cryptanalysis.
 
 We stress that our attacks could be applied in the future against today's communications, and recommend caution when choosing symmetric constructions for cases where long-term security is expected.
\end{abstract}
\section{Introduction}\label{sec:intro}
Due to Shor's algorithm~\cite{FOCS:Shor94}, quantum computing has significantly changed cryptography, despite its currently theoretical nature.

In public-key cryptography, this has led to the thriving field of quantum-safe cryptography and an ongoing competition organized by the NIST~\cite{NISTPQC} will propose new standards for key exchange and signatures. In the meantime, quantum circuits for Shor's algorithm have been proposed and improved over time~\cite{PQCRYPTO:HJNRS20,ARXIV:GidEke19,EPRINT:BBHL20,QIC:HanRoeSvo17}, leading to a better understanding of the precise resources needed for a quantum computer to be threatening.

In symmetric cryptography, it has long been thought that the only threat was the quantum acceleration on exhaustive search. This has changed with works on dedicated cryptanalysis of block ciphers~\cite{ToSC:BonNaySch19}, hash functions~\cite{EC:HosSas20}, and the many cryptanalyses that rely on Simon's algorithm~\cite{ISIT:KuwMor10,C:KLLN16,SAC:Bonnetain17,AC:LeaMay17,SAC:BonNaySch19,AC:BHNSS19}. Nevertheless, work on quantum circuits focuses mainly on exhaustive key search, and specifically on AES key search~\cite{EC:JNRV20,SAC:DavPri20,IEEE.TQE:LanPhaSte20,QIP:ASAM18,PQC:GLRS16}. Hence, many quantum attacks in symmetric cryptography are either only known asymptotically, or only with rough estimates.

\subsubsection{Our Contributions.} We present the first quantum circuits that implement the offline Simon's algorithm~\cite{AC:BHNSS19}, and propose cost estimates for the attack against the MAC \Chaskey{}, the block cipher \Prince{}, and the NIST lightweight candidate AEAD scheme \Elephant.

We stress that these attacks, as Shor's algorithm, could by applied against \emph{today}'s communications: a patient attacker could gather the required data now and wait until a powerful enough quantum computer is available to run the attack.  

Using Q\#, we designed and implemented multiple quantum circuits of independent interest: an efficient reversible circuit to solve boolean linear equations, and optimized quantum circuits for \Chaskey{}, \Prince{} and the two permutations used in \Elephant, \spongent{} and \keccak.

We find that \Prince{} and \Chaskey{} are especially vulnerable to this attack, requiring only $2^{65}$ qubit operations to recover the key. For comparison, Shor's algorithm requires $2^{31}$ similar operations to break RSA-2048. \Elephant{} suffers much less: it has a larger state size, with the same data limitation and key size. This makes the \Elephant{} cryptanalysis slightly more costly than exhaustive search.

\def\sectionautorefname{Section}
\subsubsection{Outline.} \autoref{sec:prel} presents the basics of quantum computing, the constructions we will attack and the generic quantum attacks against them.
\autoref{sec:offline-simon} presents the offline Simon's algorithm, the quantum algorithm we implement. \autoref{sec:simon-atk} presents Simon-based cryptanalysis and details for each construction the attack model and principles. In \autoref{sec:algebra}, we propose a new optimized quantum circuit to solve boolean linear equations reversibly.
\autoref{sec:circuits} presents our design of quantum circuits for the constructions we attack, as well as our optimization strategies. \autoref{sec:estimates} details the cost estimates of our attacks.

\section{Preliminaries}\label{sec:prel}
\subsection{Quantum computing}
For our purposes a quantum computer is a collection of \emph{qubits}, objects with a joint \emph{quantum state} represented by a projective complex vector space of dimension $2^n$, for $n$ qubits. We model the quantum computer as a peripheral of some classical controller~\cite{C:JaqSch2019}, which alters the quantum state by applying \emph{gates}. These interventions apply to one or more qubits, and the controller is free to apply gates simultaneously to disjoint sets of qubits. The cost of a quantum algorithm is then measured in the number of interventions applied. For quantum computers today, and for surface codes in the future, ``gates'' are not distinct physical objects, but an operation that we perform on the quantum computer. Hence, $2^{65}$ gates does not imply $2^{65}$ physical components, but it does imply performing some process $2^{65}$ times, and so we focus on the total cost of these processes. For this reason, we will often refer to gates as ``operations'' or ``qubit operations''.

The algorithms we analyze are definitively in a fault-tolerant era of quantum computing, where quantum error correction enables large computations. As surface codes are the most promising error correction candidate today~\cite{PRA:FMMC2012}, we focus on costs relevant to surface codes. We pay special attention to the number of T-gates, which are the most expensive gate on surface codes, and we do not give any extra cost to measurements.

While the attack depends on quantum interference, the most expensive subroutines are quantum emulations of classical algorithms: block ciphers, linear algebra, and memory access. Thus, we can design and test these subroutines even at cryptographic sizes. We use the Q\# programming language for this~\cite{ACM:SGA+2018}.

\begin{figure}
 \begin{subfigure}{0.23\textwidth}
 \begin{minipage}[b][2cm][b]{\textwidth}
 \centering
  \providecommand{\ket}[1]{\left\vert #1\right\rangle}
\begin{tikzpicture}[scale=1.000000,x=1pt,y=1pt]
\filldraw[color=white] (0.000000, -7.500000) rectangle (18.000000, 7.500000);
\draw[color=black] (0.000000,0.000000) -- (18.000000,0.000000);
\draw[color=black] (0.000000,0.000000) node[left] {$\ket{a}$};
\begin{scope}
\draw[fill=white] (9.000000, 0.000000) circle(3.000000pt);
\clip (9.000000, 0.000000) circle(3.000000pt);
\draw (6.000000, 0.000000) -- (12.000000, 0.000000);
\draw (9.000000, -3.000000) -- (9.000000, 3.000000);
\end{scope}
\draw[color=black] (18.000000,0.000000) node[right] {$\ket{a\oplus 1}$};
\end{tikzpicture}
  \caption{Pauli $X$ gate, or NOT gate.}
 \end{minipage}
 \end{subfigure}%
\begin{subfigure}{0.23\textwidth}
 \begin{minipage}[b][2cm][b]{\textwidth}
 \centering
  \providecommand{\ket}[1]{\left\vert #1\right\rangle}
\begin{tikzpicture}[scale=1.000000,x=1pt,y=1pt]
\filldraw[color=white] (0.000000, -7.500000) rectangle (18.000000, 22.500000);
\draw[color=black] (0.000000,15.000000) -- (18.000000,15.000000);
\draw[color=black] (0.000000,15.000000) node[left] {$\ket{a}$};
\draw[color=black] (0.000000,0.000000) -- (18.000000,0.000000);
\draw[color=black] (0.000000,0.000000) node[left] {$\ket{b}$};
\draw (9.000000,15.000000) -- (9.000000,0.000000);
\filldraw (9.000000, 15.000000) circle(1.500000pt);
\begin{scope}
\draw[fill=white] (9.000000, 0.000000) circle(3.000000pt);
\clip (9.000000, 0.000000) circle(3.000000pt);
\draw (6.000000, 0.000000) -- (12.000000, 0.000000);
\draw (9.000000, -3.000000) -- (9.000000, 3.000000);
\end{scope}
\draw[color=black] (18.000000,15.000000) node[right] {$\ket{a}$};
\draw[color=black] (18.000000,0.000000) node[right] {$\ket{a\oplus b}$};
\end{tikzpicture}
   \caption{CNOT gate}
 \end{minipage}
\end{subfigure}%
\begin{subfigure}{0.23\textwidth}
 \begin{minipage}[b][2cm][b]{\textwidth}
 \centering
  \providecommand{\ket}[1]{\left\vert #1\right\rangle}
\begin{tikzpicture}[scale=1.000000,x=1pt,y=1pt]
\filldraw[color=white] (0.000000, -7.500000) rectangle (18.000000, 37.500000);
\draw[color=black] (0.000000,30.000000) -- (18.000000,30.000000);
\draw[color=black] (0.000000,30.000000) node[left] {$\ket{a}$};
\draw[color=black] (0.000000,15.000000) -- (18.000000,15.000000);
\draw[color=black] (0.000000,15.000000) node[left] {$\ket{b}$};
\draw[color=black] (9.000000,0.000000) -- (18.000000,0.000000);
\draw (9.000000,30.000000) -- (9.000000,0.000000);
\filldraw (9.000000, 30.000000) circle(1.500000pt);
\filldraw (9.000000, 15.000000) circle(1.500000pt);
\draw[color=black] (18.000000,30.000000) node[right] {$\ket{a}$};
\draw[color=black] (18.000000,15.000000) node[right] {$\ket{b}$};
\draw[color=black] (18.000000,0.000000) node[right] {$\ket{a\wedge b}$};
\end{tikzpicture}
  \caption{AND gate}
 \end{minipage}
\end{subfigure}%
\begin{subfigure}{0.3\textwidth}
 \begin{minipage}[b][2cm][b]{\textwidth}
 \centering
  \providecommand{\ket}[1]{\left\vert #1\right\rangle}
\begin{tikzpicture}[scale=1.000000,x=1pt,y=1pt]
\filldraw[color=white] (0.000000, -7.500000) rectangle (18.000000, 37.500000);
\draw[color=black] (0.000000,30.000000) -- (18.000000,30.000000);
\draw[color=black] (0.000000,30.000000) node[left] {$\ket{a}$};
\draw[color=black] (0.000000,15.000000) -- (18.000000,15.000000);
\draw[color=black] (0.000000,15.000000) node[left] {$\ket{b}$};
\draw[color=black] (0.000000,0.000000) -- (18.000000,0.000000);
\draw[color=black] (0.000000,0.000000) node[left] {$\ket{c}$};
\draw (9.000000,30.000000) -- (9.000000,0.000000);
\filldraw (9.000000, 30.000000) circle(1.500000pt);
\filldraw (9.000000, 15.000000) circle(1.500000pt);
\begin{scope}
\draw[fill=white] (9.000000, 0.000000) circle(3.000000pt);
\clip (9.000000, 0.000000) circle(3.000000pt);
\draw (6.000000, 0.000000) -- (12.000000, 0.000000);
\draw (9.000000, -3.000000) -- (9.000000, 3.000000);
\end{scope}
\draw[color=black] (18.000000,30.000000) node[right] {$\ket{a}$};
\draw[color=black] (18.000000,15.000000) node[right] {$\ket{b}$};
\draw[color=black] (18.000000,0.000000) node[right] {$\ket{c\oplus \left(a\wedge b\right)}$};
\end{tikzpicture}
   \caption{Toffoli gate}
 \end{minipage}
\end{subfigure}%
\caption{Quantum gates used in quantum implementations of classical circuits}\label{fig:gates}
\end{figure}
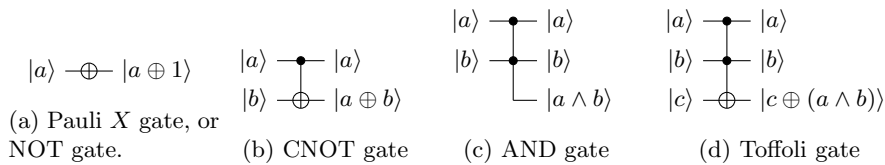

We use the Clifford+T gate set with measurements, though we design circuits using only X, CNOT, Toffoli\footnote{Confusingly, the ``T'' in ``T-gate'' does not stand for Toffoli; they are distinct gates.}, and AND operations. These operations act like classical bit operations on bitstrings, hence they are efficient to simulate. The Toffolis and ANDs are further decomposed into Clifford+T operations, and only Toffoli and AND require T operations. \autoref{fig:gates} summarizes the quantum gates we use to implement reversible classical circuits.

We did not explore any fully quantum techniques (such as measurement-based uncomputation) for these classical tasks, beyond atomic operations present in Q\#, such as measurement-based ANDs.

NIST's security levels for post-quantum cryptography emphasize the maximum circuit depth available to an adversary~\cite{NISTPQC}. Since Grover-like algorithms parallelize badly~\cite{PRA:Zalka99}, attacks that finish quickly cost much more than attacks that are allowed to take a long time. While this also affects our attack, our goal is to demonstrate another aspect of post-quantum security, rather than to compare to post-quantum asymmetric cryptography, so we do not account for depth limits. 

\subsection{Generic designs}

\subsubsection{Even-Mansour.}
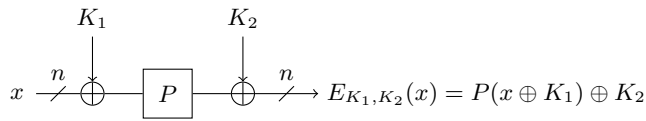
\begin{figure}[h]
\centering
\begin{tikzpicture}
  \draw
  node at (0,0)[l,name=m] {$x$}
  node [xor,circle, name=xor1, right of=m] {}
  node [block,name=f1, right of=xor1] {$P$}
  node [xor,circle, name=xor2, right of=f1] {}
  node [name=k1, above of=xor1] {$K_{1}$}
  node [name=k2, above of=xor2] {$K_{2}$}
  node [name=out, right of=xor2, right] {$E_{K_{1},K_{2}}(x) = P(x\oplus K_{1})\oplus K_{2}$};
  \draw[->] (m) -- node {\midsection} node [above=0.2em] {$n$} (xor1) -- (f1) -- (xor2) -- node {\midsection} node [above=0.2em] {$n$} (out);
  \draw[->] (k1) -- (xor1);
  \draw[->] (k2) -- (xor2);
 \end{tikzpicture}
\caption{The Even-Mansour construction. $P$ is a public permutation.}
\label{fig:EM}
\end{figure}

The Even-Mansour construction~\cite{JC:EveMan97}, presented in \autoref{fig:EM}, is a very minimal block cipher, with provable classical security: assuming $P$ has been chosen randomly, any key recovery requires an amount of time $T$ and data $D$ that satisfies $TD\geq 2^n$.

\subsubsection{FX construction.}
\begin{figure}[h]
\centering
\begin{tikzpicture}
  \draw
  node at (0,0)[l,name=m] {$x$}
  node [xor,circle, name=xor1, right of=m] {}
  node [block,name=f1, right of=xor1] {$E_K$}
  node [xor,circle, name=xor2, right of=f1] {}
  node [name=k1, above of=xor1] {$K_{1}$}
  node [name=k2, above of=xor2] {$K_{2}$};
  \path (k1) -- node[name=k] {$K$} (k2);
  \draw
  node [name=out, right of=xor2, right] {$FX_{K_{1},K,K_{2}}(x) = E_K(x\oplus K_{1})\oplus K_{2}$};
  \draw[->] (k) -- node {\midsection} node [left=0.2em] {$k$} (f1);
  \draw[->] (m)  -- node {\midsection} node [above=0.2em] {$n$} (xor1) -- (f1) -- (xor2) -- node {\midsection} node [above=0.2em] {$n$} (out);
  \draw[->] (k1) -- (xor1);
  \draw[->] (k2) -- (xor2);
 \end{tikzpicture}
\caption{The FX construction. $E_K$ is a block cipher.}
\label{fig:FX}
\end{figure}
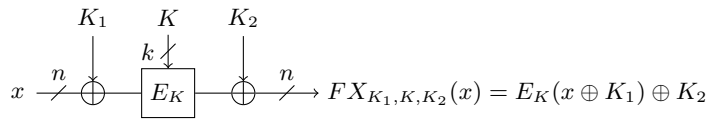

The FX construction~\cite{C:KilRog96} is a simple way to extend the key length of a block cipher: it adds two whitening keys, at the input and the output of the cipher, as presented on \autoref{fig:FX}.

\subsection{Target constructions}\label{sec:targets}

\subsubsection{Chaskey.} 
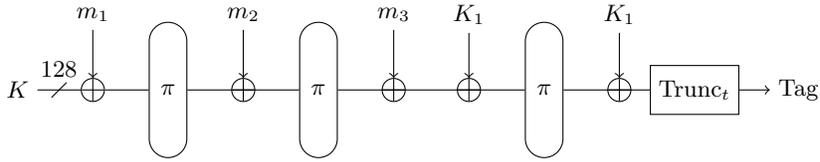
\begin{figure}[h]
  \centering
 \begin{tikzpicture}
  \draw
  node at (0,0)[l,name=m] {$K$}
  node [xor,circle, name=xor1, right of=m] {}
  node [sponge,name=f1, right of=xor1] {$\pi$}
  node [xor,circle, name=xor2, right of=f1] {}
  node [sponge,name=f2, right of=xor2] {$\pi$}
  node [xor,circle, name=xor3, right of=f2] {}
  node [xor,circle, name=xor4, right of=xor3] {}
    node [sponge,name=f3, right of=xor4] {$\pi$}
  node [xor,circle, name=xor5, right of=f3] {}
  node [name=k1, above of=xor1] {$m_1$}
  node [name=k2, above of=xor2] {$m_2$}
  node [name=k3, above of=xor3] {$m_3$}
  node [name=k4, above of=xor4] {$K_1$}
  node [name=k5, above of=xor5] {$K_1$}
  node [block, name=trunk, right of=xor5] {$\text{Trunc}_t$}
  node [name=out, right of=trunk, right] {Tag};
  \draw[->] (m) -- node {\midsection} node [above=0.2em] {$128$} (xor1) -- (f1) -- (f2) -- (f3) -- (trunk) -- (out);
  \draw[->] (k1) -- (xor1);
  \draw[->] (k2) -- (xor2);
  \draw[->] (k3) -- (xor3);
  \draw[->] (k4) -- (xor4);
  \draw[->] (k5) -- (xor5);
 \end{tikzpicture}
\caption{Chaskey mode for a message of 3 blocks.}
\label{fig:chaskey}
\end{figure}

\Chaskey~\cite{SAC:MMVWPV14} is a lightweight MAC oriented to 32-bits architectures. It uses a mode that can be seen as a combination of Even-Mansour and CBC-MAC, described in \autoref{fig:chaskey}, with a 128-bit ARX permutation $\pi$.

It uses a 128-bit key $K$, from which the key $K_1$ is derived: $K_1 = 2K$, with a multiplication in the finite field $\mathbb{F}_2[X]/\left(X^{128}+X^7+X^2+X+1\right)$.

It outputs a $t$-bit tag, with $t\leq 128$ specified by the user. In the original design, the permutation contained 8 rounds. As the 7-rounds permutation happened to be broken~\cite{EC:Leurent16}, \Chaskey{} with a 12-rounds permutation is included in the standard ISO/IEC 29192-6~\cite{ISO:Chaskey}.

\Chaskey{} has a data limitation of $2^{48}$ message blocks with the same key, which corresponds to $2^{55}$ bits.

\paragraph{Classical security.} Because of the Even-Mansour construct, \Chaskey{} can be attacked with a time-data tradeoff that satisfies $TD \geq 2^{128}$, which is why the data is limited to $2^{48}$ blocks.

\subsubsection{PRINCE.} 
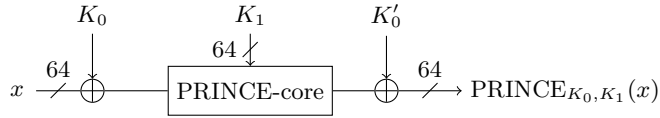
\begin{figure}[h]
\centering
\begin{tikzpicture}
  \draw
  node at (0,0)[l,name=m] {$x$}
  node [xor,circle, name=xor1, right of=m] {}
  node [block,name=f1, right of=xor1,right] {PRINCE-core}
  node [xor,circle, name=xor2, right of=f1,right=2em] {}
  node [name=k1, above of=xor1] {$K_{0}$}
  node [name=k2, above of=xor2] {$K'_0$}
  node (k) at (k1-|f1) {$K_1$};
  \draw
  node [name=out, right of=xor2, right] {PRINCE$_{K_0,K_1}(x)$};
  \draw[->] (k) -- node {\midsection} node [left=0.2em] {$64$} (f1);
  \draw[->] (m)  -- node {\midsection} node [above=0.2em] {$64$} (xor1) -- (f1) -- (xor2) -- node {\midsection} node [above=0.2em] {$64$} (out);
  \draw[->] (k1) -- (xor1);
  \draw[->] (k2) -- (xor2);
 \end{tikzpicture}
\caption{The \Prince{} cipher. $K'_0 = (K_{0} \ggg 1)\oplus (K_{0} \gg 63)$.}
\label{fig:prince}
\end{figure}

\Prince~\cite{AC:BCGKKK12} is a low-latency block cipher, with a 64 bit block size and a 128 bit key, split into two 64-bit keys, $K_0$ and $K_1$. It follows the FX construction, as presented in \autoref{fig:prince}.

Notably, some microcontrollers use \Prince{} to encrypt memory~\cite{TECH:Prince}.

\paragraph{Classical security.} \Prince{} claims a data-time tradeoff of $TD\geq 2^{126}$. It has been analyzed extensively~\cite{FSE:JNPWW13,FSE:SBYWNZ13,AC:FouJouMav14,FSE:CFGNR14,EC:Dinur15,FSE:DerPer15,AFRICACRYPT:RasRad16,INDOCRYPT:GraRec16}, and so far the claim holds.

Very recently, a new version of \Prince{}, \Prince{}v2~\cite{SAC:BEKLLMNRTW20} was proposed. While this new version is very close to \Prince{}, it does not have the FX structure, and each round uses alternatively $K_0$ or $K_1$. This makes \Prince{}v2 immune to the attack we present here.

\subsubsection{Elephant.} \Elephant~\cite{Elephant} is an authenticated encryption with associated data (AEAD) scheme, and a 2nd-round candidate in the NIST lightweight authenticated encryption competition~\cite{NISTLW}. It is a block-oriented construction whose encryption shares some similarities with the counter mode, with an encrypt-then-MAC authentication.

\Elephant{} uses a 128-bit key $K$ and a 96-bit nonce $N$. It comes in 3 variants, with a different permutation $P$ and a different security level:
\begin{description}
 \item[\Elephant-160] uses the 160-bit permutation \spongent-$\pi$[160]~\cite{CHES:BKLTVV11}. Its expected classical security is $2^{112}$ with data limited to $2^{53}$ bits processed.
 \item[\Elephant-176] uses the 176-bit permutation \spongent-$\pi$[176]~\cite{CHES:BKLTVV11}. Its expected classical security is $2^{127}$ with a data limited to $2^{53}$ bits processed.
 \item[\Elephant-200] uses the 200-bit permutation \keccak-$f$[200]~\cite{Keccak}. Its expected classical security is $2^{127}$ with a data limited to $2^{77}$ bits processed.
\end{description}

\begin{figure}
 \centering
 \begin{tikzpicture}[auto, node distance=1cm,inner sep=0em ]
  \draw
  node at (0,0) [l,name=w0] {$N||0$}
  node [xor,name=xor0,below of=w0] {}
  node [block,name=ek0, below of=xor0] {$P$}
  node [left of=ek0,node distance=1cm,name=t0,left] {mask$^{0,0}_K$};
  \path (t0) -- node (t1) {} (ek0);
  \draw (w0) -- (xor0) -- (ek0);
  \draw
  node [xor, name=xor1,below of=ek0] {}
  node [xor, name=xor2,below of=xor1] {}
  node [left of=xor2, node distance=1cm, name=t2] {$M_1$}
  node [l,below of=xor2,name=c0] {$C_1$};
  \draw[->] (t0) -| (t1.center) |- (xor0);
  \draw[->] (t1.center) |- (xor1);
  \draw[->] (t2) -- (xor2);
  \draw[->] (ek0) -- (xor1) -- (xor2) -- (c0);
  \draw node[right of=ek0, node distance=1cm] {\dots};
  \draw
  node [l,right of=w0, node distance=4cm,name=w0] {$N||0$}
  node [xor,name=xor0,below of=w0] {}
  node [block,name=ek0, below of=xor0] {$P$}
  node [left of=ek0,node distance=1cm,name=t0,left] {mask$^{\ell-1,0}_K$};
  \path (t0) -- node (t1) {} (ek0);
  \draw (w0) -- (xor0) -- (ek0);
  \draw
  node [xor, name=xor1,below of=ek0] {}
  node [xor, name=xor2,below of=xor1] {}
  node [left of=xor2, node distance=1cm, name=t2] {$M_\ell$}
  node [l,below of=xor2,name=c0] {$C_\ell$};
  \draw[->] (t0) -| (t1.center) |- (xor0);
  \draw[->] (t1.center) |- (xor1);
  \draw[->] (t2) -- (xor2);
  \draw[->] (ek0) -- (xor1) -- (xor2) -- (c0);
  
 \end{tikzpicture}
\caption{\Elephant{} encryption of the message $(M_i)$.}\label{fig:elephant}
\end{figure}
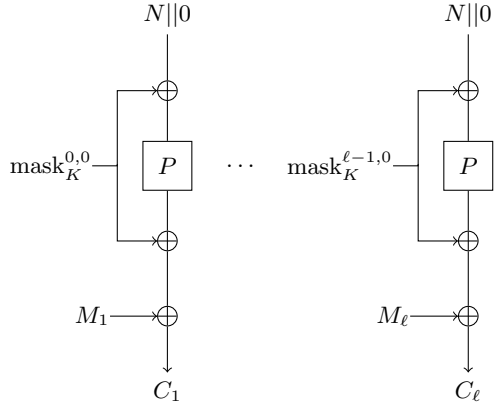

The encryption of a message is presented on \autoref{fig:elephant}. The mask values are computed from the expanded key $K' = P(K||0)$, and two LFSR $\phi_a$ and $\phi_b$:
\[
 \text{mask}^{i,j}_K = \phi_b^{(j)}\circ\phi_a^{(i)}(K')
\]
For encryption, only $j = 0$ is used. Masks with $j = 1$ and $j = 2$ are used to compute the tag.

A new version of \Elephant, \Elephant{} v2~\cite{Elephant-update}, has been proposed for the  third round of the NIST lightweight competition. There are only two differences between the versions: the encryption uses masks with $j = 1$ for encryption, and the tag computation is different. This does not affect our attack.

\subsection{Generic attacks}
There are two types of attacks that can always be applied on the structures we're attacking.

\subsubsection{Key search.} As the constructions contain some secret material, it is possible to brute-force it. Classically, this will cost $2^k$ computations of the construction.

Its quantum equivalent uses amplitude amplification~\cite{BHMT02} to recover the key, and requires $\frac\pi22^{k/2}$ computations of the construction, assuming one computation can uniquely identify the key.

\subsubsection{Collision finding.} The Even-Mansour construction can be attacked by looking for collisions~\cite{EC:DunKelSha12}:
let's consider that we have queried $2^d$ Even-Mansour encryptions. For any $\delta$, we can compute a list of elements of the form
\[
 E_{K_1,K_2}(x) \oplus P(x\oplus \delta) = P(x \oplus K_1)\oplus P(x\oplus \delta) \oplus K_2
\]

If the list happens to contain two messages $x,y$ such that $x\oplus y\oplus \delta = K_1$, then we have
 $P(x\oplus \delta) = P(y\oplus K_1)$ and conversely $P(y\oplus \delta) = P(x\oplus K_1)$. Hence, the list will contain a collision.
 
 As the list is of size $2^d$, this will occur with probability $2^{2d-n}$, which means we need to try $2^{n-2d}$ distinct $\delta$. Overall, as one try costs $2^d$, the total time cost is $T = 2^n/2^d$, with $2^d$ data, for a tradeoff of $DT = 2^n$.

\paragraph{Quantum version.} There are multiple quantum algorithms to compute collisions. The most well known matches the query lower bound of $\Omega\left(2^{n/3}\right)$~\cite{BHT98}. It however requires the QRAM model, and there is no known time-efficient implementation of this algorithm.

More recently, a quantum algorithm based on distinguished points has been proposed~\cite{AC:ChaNaySch17}, with a time cost in $\mathcal{O}\left(2^{2n/5}\right)$ or $\mathcal{O}\left(2^{3n/7}\right)$, depending whether one of the colliding functions can be queried quantumly or not. This algorithm was used in~\cite{RSA:HosSas18} to propose quantum attacks on Even-Mansour with the tradeoff $DT^6 = 2^{3n}$.

\paragraph{Collisions for FX.} The FX construction can be attacked simply by checking wether or not the Even-Mansour attack works given an inner key guess. This changes the tradeoffs, replacing $n$ with $n+k$.

\begin{remark}
 One may consider that searching for the key will always be more expensive than looking for collisions. This is not always the case: collision-finding depends on the state size, and key search on the key size (though the two are often equal).
\end{remark}

\begin{remark} 
The classical security claims of our target constructions match the tradeoff $DT = 2^n$ or $DT = 2^{n+k}$.
\end{remark}

\section{The offline Simon's algorithm}\label{sec:offline-simon}

The following sections present the algorithmic core of our attacks, which amounts to finding a periodic function.
\begin{definition}[Periodic function]
 Let $f~: \zo^n \rightarrow X$ be a function. $f$ is periodic if there exists an $s$ such that for all $x$, $f(x) = f (x\oplus s)$.
\end{definition}

From an abstract point of view, our attacks can be seen as instances of the following problem:
\begin{problem}[Offline Simon's problem]\label{problem:offline}
Let $f~: \zo^k \times \zo^n \rightarrow \zo^m$ and $E~:\zo^n \rightarrow \zo^m$ be functions, with $s\in \zo^n, c \in \zo^m$  such that there exist a unique $i_0 \in \zo^m$ such that $E(x) = f(i_0,x\oplus s)\oplus c$. Find $i_0$ and $s$. 
\end{problem}
Solving this problem reduces to finding a periodic function, as the function $E(x)\oplus f(i_0,x)$ has period $s$.
Here, $E$ will be a secret function (a block cipher, for example) that we can only query classically, and $f$ will be computable quantumly.

\subsection{Simon's algorithm} Simon's algorithm~\cite{FOCS:Simon94} solves the following problem in polynomial time:
\begin{problem}[Simon's Problem]\label{problem:simon}
 Let $n$ be an integer and $X$ a set.
 Let $f : \{0,1\}^n \to X$ be a function such that for all $(x,y) \in \left(\{0,1\}^n\right)^2$ with $x\neq y$,
 $[f(x) = f(y) \Leftrightarrow x = y \oplus s]$. Given oracle access to $f$, find $s$.
\end{problem}

It does so using \autoref{fig:simon}, which is described as \autoref{alg:simon-routine}.

\begin{circuit}[h]
\centering
    \begin{tikzpicture}
  \draw 
  node at (0,0) (in) {$\{0,1\}^n : \ket{0}$}
  node [below of=in] (anc) {$\{0,1\}^n : \ket{0}$}

  node [right=of in, block] (qft1) {$H$}
  node [below of=qft1] (b1) {};
  \path (qft1) -- node (middle) {} (b1);
  \draw
  node [block,right=of middle, minimum height=6em] (f){$O_f$}
  node [right= of qft1] (f1) {}
  node [right= of b1] (f2) {}
  
  node [right=of f1.center, block] (qft) {$H$}
  node [below of=qft] (b) {}
  
  node [right=of qft.center] (ctrlo) {$\measure : j$}
  node [right=of b.center] (anco) {$\measure : f(x_0) = f(x_0\oplus s)$}
;

  \draw (in) -- (qft1) -- (f.west|-qft1) (f.east|-qft) -- (qft) -- (ctrlo)
        (anc) -- (f.west|-anc) (f.east|-anco) -- (anco)
  ;
 \end{tikzpicture}
 \caption{Simon's circuit}\label{fig:simon}
\end{circuit}

\begin{algorithm}
\begin{algorithmic}[1]
  \Statex \textbf{Input:} $n$, $O_f : \ket{x}\ket{0} \mapsto \ket{x}\ket{f(x)}$ with $f :\zo^n \to X$ a Simon function
  \Statex \textbf{Output:} $j$ with $j\cdot s = 0$
  \State Initialize two n-bits registers : $\ket0\ket0$
  \State Apply $H$ gates on the first register, to compute $\sum_{x = 0}^{2^{n}-1} \ket{x}\ket0$
  \State Apply $O_f$, to compute $\sum_{x = 0}^{2^{n}-1}\ket{x}\ket{f(x)}$\label{step:io}
  \State Reapply $H$ gates on the register, to compute \[\sum_{x = 0}^{2^n-1}\sum_{j = 0}^{2^n-1}(-1)^{x\cdot j}\ket{j}\ket{f(x)}\]\label{step:h2}
  \State We can factor the $x$ that have the same $f(x)$, and rewrite the state as
  \[\sum_{x\in \zo^n/(s)}\sum_{j = 0}^{2^n-1}\left((-1)^{x\cdot j}+(-1)^{(x\oplus s)\cdot j}\right)\ket{j}\ket{f(x)} \]
  \State Measure $j, f(x)$, return them.
\end{algorithmic}
\caption{Simon's routine}\label{alg:simon-routine}
\end{algorithm}

Now, from \autoref{alg:simon-routine}, we see that the $j$ we can measure must fulfill $(-1)^{x\cdot j}+(-1)^{(x\oplus s)\cdot j} \neq 0$, that is, $s \cdot j = 0$. Hence, this routine can only produce values orthogonal to the secret.

\begin{remark} If the function is not periodic, then random values will be measured, and the set of values can be of rank $n$.
\end{remark}

\subsubsection{Full algorithm.} From this circuit, we recover the complete value of $s$ by obtaining $\mathcal{O}(n)$ queries, and using linear algebra classically to compute $s$.

\subsubsection{Reversible implementations of Simon's algorithm.} Without the final measurement,  \autoref{alg:simon-routine} becomes a reversible quantum circuit that computes in its first register the uniform superposition of values orthogonal to $s$. Hence, if we apply it multiple times in parallel, we can reversibly compute the value of $s$, assuming we also have a quantum circuit for the linear algebra. We present such a circuit in \autoref{sec:algebra}.

\subsubsection{Simon's algorithm as a distinguisher.} As Simon's algorithm can compute a period, it can also determine wether a given function is periodic or not. With enough sampled vectors, their rank will be at most $n-1$ if the function is periodic, and will likely be $n$ if the function is not. This principle can be used in quantum distinguishers. 

\subsection{Grover-meets-Simon} The Grover-meets-Simon algorithm~\cite{AC:LeaMay17} performs a quantum search that uses Simon's algorithm to identify the correct guess. This is possible as Simon's algorithm can be implemented reversibly. Grover-meets-Simon solves the following problem:
\begin{problem}[Search for a periodic function]\label{problem:gms}
  Let $n$ be an integer and $X$ a set.
 Let $f : \{0,1\}^k\times\{0,1\}^n \to X$ be a function such that there exists a unique $i_0$ such that $f(i_0,\cdot)$ is periodic. Find $i_0$ and the period of $f(i_0,\cdot)$.
\end{problem}

\autoref{alg:gms} solves this problem by simply testing wether or not the function $f(i,\cdot)$ is periodic, using Simon's algorithm as in \autoref{fig:simon-gms}.

This algorithm has a cost of $\mathcal{O}\left(n2^{k/2}\right)$ queries and $\mathcal{O}\left(n^32^{k/2}\right)$ time, as each iteration of the quantum search requires an application of Simon's algorithm, which needs $\mathcal{O}\left(n\right)$ queries plus $\mathcal{O}\left(n^3\right)$ for the linear algebra.

\begin{algorithm}
\begin{algorithmic}[1]
 \AAmp{$i\in \{0,1\}^k$}
   \State Apply Simon's algorithm on $f(i,\cdot)$
   \State {$b\gets$ the period is not 0}\Comment{Vector set of rank $<n$}
   \If{ $b$ }
   \State Do a phase shift
   \EndIf
   \State Uncompute Simon's algorithm
 \EndAA
\end{algorithmic}
\caption{Grover-meets-Simon algorithm~\cite{AC:LeaMay17}}\label{alg:gms}
\end{algorithm}

\begin{circuit}[h]
\centering
\begin{tikzpicture}
  \draw 
  node at (0,0) (x0) {$\ket{0}$}
  node [above of=x0] (i) {$\ket{i}$}
  node [below of=x0, above] (y0) {$\ket{0}$}
  node [right of=x0, block] (qft0) {$H$}
  node [right of=y0] (b1) {};
  \path (qft0) -- node (middle) {} (b1);
  \draw
  node [block,right of=middle, minimum height=4em] (f0){$f$}
  node [right of=qft0] (f) {}
  node (g1) at (f-|f0.east) {}
  node [right of=g1, block] (Qft0) {$H$}
  node[right of=Qft0] (q) {};
  node [below of=Qft0] (b) {};
  \draw (x0) -- (qft0) -- (f0.west|-qft0) (f0.east|-Qft0) -- (Qft0) -- (q|-Qft0)  ;
  \draw (y0) -- (f0.west|-y0) (f0.east|-y0) -- (Qft0.center|-y0);
    \draw (i) -- (i-|Qft0);

  \draw 
  node [below of=y0] (x0) {$\ket{0}$}
  node [below of=x0, above] (y0) {$\ket{0}$}
  node [right of=x0, block] (qft0) {$H$}
  node [right of=y0] (b1) {};
  \path (qft0) -- node (middle) {} (b1);
  \draw
  node [block,right of=middle, minimum height=4em] (f1){$f$}
  node [right of=qft0] (f) {}
  node (g1) at (f-|f1.east) {}
  node [right of=g1, block] (Qft0) {$H$}
  node [below of=Qft0] (b) {};
  \draw (x0) -- (qft0) -- (f1.west|-qft0) (f1.east|-Qft0) -- (Qft0)-- (q|-Qft0)  ;
  \draw (y0) -- (f1.west|-y0) (f1.east|-y0) -- (Qft0.center|-y0);
    
  \draw 
  node [below of=y0, below=0.3cm] (x0) {$\ket{0}$}
  node [below of=x0, above] (y0) {$\ket{0}$}
  node [right of=x0, block] (qft0) {$H$}
  node [right of=y0] (b1) {};
  \path (qft0) -- node (middle) {} (b1);
  \draw
  node [block,right of=middle, minimum height=4em] (f2){$f$}
  node [right of=qft0] (f) {}
  node (g1) at (f-|f2.east) {}
  node [right of=g1, block] (Qft0) {$H$}
  node [below of=Qft0] (b) {};
  \draw (x0) -- (qft0) -- (f2.west|-qft0) (f2.east|-Qft0) -- (Qft0) -- (q|-Qft0) ;
  \draw (y0) -- (f2.west|-y0) (f2.east|-y0) -- (Qft0.center|-y0);
  \path (f1) -- node (m) {\dots} (f2);
  \draw (i) -| (f0) -- (f1) --(m) --(f2);
  \path (f0) -- node (m) {} (f2);
  \node[right,draw,minimum height=6cm] (ln) at (m-|q) {Linear algebra};
  \node[right=of ln] (out) {$\ket{\text{Rank} \overset?=n}$};
  \draw (ln) -- (out);
  \node[below=of ln.south, above] {Note: Ancilla qubits and unused outputs are not represented.};
\end{tikzpicture}
 \caption{Simon's circuit in Grover-meets-Simon}\label{fig:simon-gms}
\end{circuit}

\subsection{The offline Simon's algorithm} We can see \autoref{problem:offline}, the Offline Simon's problem, as a special case of \autoref{problem:gms}, a search for a periodic function, and solve it with \autoref{alg:gms}. Indeed, if we have $E(x) = f(i_0,x\oplus s)\oplus c$, then the function $E(x)\oplus f(i,x)$ will be periodic if and only if $i = i_0$, and its period will be $s$. The main limitation of this approach is that we need quantum query access to the periodic function, which is not possible if the function $E$ is only accessible classically.

The offline Simon's algorithm~\cite{AC:BHNSS19} proposes two improvements over the Grover-meets-Simon algorithm to overcome this restriction.

\subsubsection{Reusing quantum queries.} The first improvement comes from the fact that the periodic function, $E(x)\oplus f(i,x)$, has a very specific two-part structure, where the function $E(x)$ is independent of $i$. This means each occurence of the Simon test makes the exact same query to $E$. This allows a slightly different approach for the Simon test: the queries to $E$ are done once at the beginning of the procedure, and then reused for each test, as shown in \autoref{alg:off}, which uses \autoref{fig:simon-offline} instead of \autoref{fig:simon-gms}.

\def\offcolor{blue}
\begin{circuit}[h]
\centering
\begin{tikzpicture}
  \draw
  node[color=\offcolor] at (0,0) (x0) {$\ket{0}$}
  node at (-0.4,0.5) (up) {}
  node [above of=x0] (i) {$\ket{i}$}
  node [color=\offcolor,below of=x0, above] (y0) {$\ket{0}$}
  node [color=\offcolor,right of=x0, block] (qft0) {$H$}
  node [right of=y0] (b1) {};
  \path (qft0.center) -- node (middle) {} (b1.center);
  \draw
  node [color=\offcolor,block,right of=middle, minimum height=4em] (e1){$E$}
  node [right of=qft0] (e) {}
  node [block,right of=e1, right, minimum height=4em] (f0){$f$}
  node [right of=e] (f) {}
  node (g1) at (f-|f0.east) {}
  node [right of=g1, block] (Qft0) {$H$}
  node[right of=Qft0] (q) {};
  node [below of=Qft0] (b) {};
  \draw[color=\offcolor] (x0) -- (qft0) -- (e1.west|-qft0)  (e1.east|-Qft0) -- (f0.west|-qft0) ;
  \draw (f0.east|-Qft0) -- (Qft0) -- (q|-Qft0)  ;
  \draw[color=\offcolor] (y0) -- (e1.west|-y0) (e1.east|-y0) -- (f0.west|-y0);
  \draw (f0.east|-y0) -- (Qft0.center|-y0);
  \draw (i) -- (i-|Qft0);

  \draw 
  node [color=\offcolor,below of=y0] (x0) {$\ket{0}$}
  node [color=\offcolor,below of=x0, above] (y0) {$\ket{0}$}
  node [color=\offcolor,right of=x0, block] (qft0) {$H$}
  node [right of=y0] (b1) {};
  \path (qft0.center) -- node (middle) {} (b1.center);
  \draw
  node [color=\offcolor,block,right of=middle, minimum height=4em] (e1){$E$}
  node [right of=qft0] (e) {}
  node [block,right of=e1, right,minimum height=4em] (f1){$f$}
  node [right of=e] (f) {}
  node (g1) at (f-|f1.east) {}
  node [right of=g1, block] (Qft0) {$H$}
  node [below of=Qft0] (b) {};
  \draw[color=\offcolor] (x0) -- (qft0) -- (e1.west|-qft0) (e1.east|-Qft0) -- (f1.west|-qft0);
  \draw (f1.east|-Qft0) -- (Qft0)-- (q|-Qft0)  ;
  \draw[color=\offcolor] (y0) -- (e1.west|-y0) (e1.east|-y0) -- (f1.west|-y0);
  \draw (f1.east|-y0) -- (Qft0.center|-y0);
  \node[color=\offcolor,below of=qft0, below] {\vdots};
  \draw 
  node [color=\offcolor,below of=y0, below=0.3cm] (x0) {$\ket{0}$}
  node [color=\offcolor,below of=x0, above] (y0) {$\ket{0}$}
  node [color=\offcolor,right of=x0, block] (qft0) {$H$}
  node [right of=y0] (b1) {};
  \path (qft0.center) -- node (middle) {} (b1.center);
  \draw
  node [color=\offcolor,block,right of=middle, minimum height=4em] (e1){$E$}
  node [right of=qft0] (e) {}
  node [block,right of=e1, right,minimum height=4em] (f2){$f$}
  node [right of=e] (f) {}
  node (g1) at (f-|f2.east) {}
  node [right of=g1, block] (Qft0) {$H$}
  node [below of=Qft0] (b) {};
  \draw[color=\offcolor] (x0) -- (qft0) -- (e1.west|-qft0) (e1.east|-Qft0) -- (f2.west|-qft0);
  \draw (f2.east|-Qft0) -- (Qft0) -- (q|-Qft0) ;
  \draw[color=\offcolor] (y0) -- (e1.west|-y0) (e1.east|-y0) -- (f2.west|-y0);
  \draw (f2.east|-y0) -- (Qft0.center|-y0);
  \draw (i) -| (f0) -- (f1);
  \draw[dashed] (f1)--(f2);
  \path (f0) -- node (m) {} (f2);
  \node[right,draw,minimum height=6cm] (ln) at (m-|q) {Linear algebra};
  \node[right=of ln] (out) {$\ket{\text{Rank} \overset?=n}$};
  \node (down) at ($(e1.south east)+(0.3,-0.3)$) {};
  \draw[dashed,color=\offcolor] (up) rectangle (down);
  \path (up|-down) -- node (bot) {} (down);
  \node[color=\offcolor,below] at (bot) {Computed once beforehand};
  \draw (ln) -- (out);
  \node[below=of f2.south east] {Note: Ancilla qubits and unused outputs are not represented.};
\end{tikzpicture}
 \caption{Simon Circuit in the offline Simon's algorithm}\label{fig:simon-offline}
\end{circuit}

\begin{algorithm}
\begin{algorithmic}[1]
  \State Query $m$ times $E$, to compute
  \[
   \ket{\psi^m} = \bigotimes_{j=1}^m\sum_x\ket{x}\ket{E(x)}
  \]

 \AAmp{$i\in \{0,1\}^k$}
   \State From $\ket{\psi^m}$, compute $m$ times
   \[\sum_x\ket{x}\ket{E(x)\oplus f(i,x)}
   \]
   \State Apply $H$ on the input registers
   \State Compute the rank of the values in the input registers
   \If{ the rank is lower than $n$}
   \State Do a phase shift
   \EndIf
   \State Uncompute steps 5 to 3.
    \EndAA
\end{algorithmic}
\caption{The Offline Simon's algorithm~\cite{AC:BHNSS19}}\label{alg:off}
\end{algorithm}

This new approach reduces the number of quantum queries to $E$ from exponential to polynomial.

\subsubsection{Using classical queries.} The second improvement computes the states
\[
 \sum_x\ket{x}\ket{E(x)}
\]
from classical queries. We can do this if we know \emph{all} the values of $E(x)$. In that case, computing the superposition corresponds to making a QRAM query to the classical values. Because we are in the circuit model, this costs $2^n$ classical queries and $\mathcal{O}\left(2^n\right)$ quantum computations. We use an optimized circuit from~\cite{PRX:BGB+2018}.

\subsection{Simon's algorithm with additional collisions and concrete estimates}
In practice, the promise of Simon's algorithm is only partially fulfilled: for the periodic functions we consider, we can have $f(x) = f(y)$ and $x \neq y\oplus s$. This impacts Simon's algorithm, but~\cite{EPRINT:Bonnetain20} shows that for almost all functions, the cost overhead is negligible, via the following theorem:

\begin{theorem}[{\cite[Theorem 14]{EPRINT:Bonnetain20}}]\label{thm:off-rand}
 Assume that $m \geq \log_2(4e(n+k+\alpha+1))$ and $k \geq 7$. The fraction of functions in $\zo^k\times \zo^n \to \zo^m$ such that the offline Simon's algorithm, repeating $\frac{\pi}{4\arcsin{\sqrt{2^{-k}}}}$ iterations with $n+k+\alpha+1$ queries per iteration, succeeds with probability lower
 than \[
       1 - 2^{-\alpha} -\left(2^{-\alpha/2+1} +2^{-\alpha} +2^{-k/2+1}\right)^2,
      \]
is lower than $2^{n + k - \frac{2^n}{4(n+k+\alpha+1)}}$.
\end{theorem}

\autoref{thm:off-rand} tells us that Simon's algorithm needs only $(n+k+\alpha+1)$ queries, and it allows us to use functions with a small output size, which roughly halves the required number of qubits and slightly reduces the computational cost of $f$. This approach shares some similarities with the oracle compression technique from
~\cite{ARXIV:MaySch20}. We however do not consider a random set of functions applied to the output, but a carefully chosen function such that the overall computational cost is minimized.

\section{Quantum Simon-based attacks}\label{sec:simon-atk}
Since the seminal Simon-based distinguisher on the 3-round Feistel construction of Kuwakado and Morii~\cite{ISIT:KuwMor10}, many attacks that use Simon's algorithm have been proposed.  We present here the Simon-based attacks on the Even-Mansour and FX constructions, and detail how we instantiate them for the primitives presented in \autoref{sec:targets}.

\subsection{Attack on Even-Mansour} 

For Even-Mansour constructions, we can consider the function
\[
 E_{K_{1},K_{2}}(x)\oplus P(x) = P(x)\oplus P(x\oplus K_{1})\oplus K_{2}\enspace,
\]
which has period $K_1$. Hence, with access to quantum queries, Simon's algorithm can recover $K_1$ in polynomial time, from which it is trivial to recover $K_2$. This was proposed in~\cite{KuwMor12}.

\subsection{Attack on the FX construction}

The quantum attack against the FX construction proposed in~\cite{AC:LeaMay17} is based on a simple idea:
if the key is known, then this reduces to an Even-Mansour, and the previous attack applies. 
In more details, the function
\[
 FX_{K_{1},K,K_{2}}(x)\oplus E_i(x) = E_i(x)\oplus E_K(x\oplus K_{1})\oplus K_{2}
\]
has period $K_1$ if and only if $i = K$. Hence, with quantum query access, we can apply the Grover-meets-Simon algorithm to recover $K$ and $K_1$ in time $\mathcal{O}\left(2^{k/2}\right)$ if $|K| = k$.

\subsection{Offline version}
The previous attacks can be adapted to classical-query attacks thanks to the offline Simon's algorithm, as proposed in~\cite{AC:BHNSS19}.

\subsubsection{Offline attack on the FX construction.}
The periodic function of the FX construction directly fits the structure of \autoref{problem:offline}, with $E = FX_{K_{1},K,K_{2}}$ and $f(i,x) = E_i(x)$. Hence, we can attack the FX construction on a block cipher of $n$ bits with a $k$-bit key in $2^n$ classical queries and time $\mathcal{O}\left(\max\left(2^n,2^{k/2}\right)\right)$.

\subsubsection{Offline attack on Even-Mansour.} We cannot directly apply the previous attack, as it would require $2^n$ classical queries. However, if we fix $n-u$ bits in the input of the cipher, we can still obtain a periodic function:
\[
 E_{K_1,K_2}(x||0^{n-u})\oplus P(x||y) = P(x||y) \oplus P(x\oplus K_1^{1}||K_1^2)\oplus K_2
\]
with $K_1^1$ the first $n-u$ bits of $K_1$, and $K_1^2$ its last $u$ bits. This function is periodic if and only if $y = K_1^2$. Hence, we can apply the offline Simon's algorithm, at a cost of $\mathcal{O}\left(2^{u}\right)$ classical queries, and $\mathcal{O}\left(\max\left(2^{u},2^{(n-u)/2}\right)\right)$ quantum time. In this case we can choose $u$, and the cost will be minimal for $u\sim n/3$.

\begin{remark}[Truncation, affine spaces]
 Technically, the input is not required to be of the form $(x||0^{n-u})$. The attack can work with any $u$-dimensional affine space. In particular, for any fixed $c$, we can take all the inputs of the form $(x||c)$.
\end{remark}

\begin{remark}[Truncation for the FX attack]
We can also apply this input truncation technique to the FX attack. This can balance the costs if $n > k/2$.
\end{remark}

\subsubsection{Concrete estimates.} We rely on \autoref{thm:off-rand} for concrete query estimates.
We chose $\alpha = 9$, as this will ensure a success probability of around 99\%. In all the instances we consider, we have $n+k \leq 200$. Hence, an output size of $m = 11$ bits will be sufficient for our purposes.

\subsection{Attack on Chaskey}

We attack \Chaskey{} with a one-block message, which degenerates into a truncated Even-Mansour:
\[
 \text{Chaskey}(m_1) = \text{Trunc$_t$}\left(\pi(m_1\oplus K \oplus K_1) \oplus K_1\right)
\]

From~\autoref{thm:off-rand} the attack does not require the full output, so the truncation is not an issue. However, for some of the circuit optimizations in \autoref{sec:chaskey-circuit}, we assume $t\geq 96$.

We can directly apply the Even-Mansour offline attack. We do a chosen-plaintext attack, and query classically the MAC of the $2^u$ 128-bit messages of the form $0^{n-u}*$. 

Then the quantum attack recovers the value of $K \oplus K_1$. As $K_1 = 2K$, we have $K\oplus K_1 = 3K$. Thus, we can divide by 3 in the finite field to recover the key $K$, which is the master key.

\subsection{Attack on PRINCE}

We can directly apply the FX attack to \Prince{}. We do a chosen-plaintext attack, and classically query the encryption of $2^u$ 64-bit messages of the form $0^{n-u}*$. Then the quantum attack recovers $K_0$ and $K_1$, which correspond to the full \Prince{} key.

\subsection{Attack on Elephant}

To attack \Elephant, we consider the encryption of a single-block message:
\[
 E_K(M) = P\left((N||0)\oplus \text{mask}^{0,0}_K \right)\oplus \text{mask}^{0,0}_K\oplus M \enspace.
\]

This is an Even-Mansour, but the input is the nonce, not the message. Hence, with only known plaintexts, we can gain access to the values we need. To make the attack work, we need to have a set of $2^u$ nonces that form an affine space. This is no obstacle to the attack, since \Elephant{}'s security proofs assume the adversary can choose nonces as long as they do not repeat. Interestingly, if the adversary has no control of the nonces but the nonce is incremented between each query, then the nonces will still from an affine space and the attack will go through.

As we have an Even-Mansour construction, we can apply the offline Simon attack, which will recover the value of $\text{mask}^{0,0}_K = K' = P(K||0)$. This expanded key is sufficient to compute all the masks in \Elephant. Moreover, as $P$ is a permutation, we can also recover the 128-bit master key $K$.

\section{A quantum circuit to solve boolean linear equations}\label{sec:algebra}

 In this section, we present a quantum algorithm that can compute, given $m$ $n$-bit vectors as input, the rank of their span or a basis of its dual. At its core, it uses \autoref{alg:algebra}, which computes a basis of the span in triangular form.  From this we can easily compute the rank or any orthogonal vector.
 
\begin{figure}
\centering
\begin{tikzpicture}
  \node (r1) at (0,0) [draw,rectangle,minimum width=6em] {}
  node (x1) [right of=r1, right] {$x_1[1..n]$}
  node (v1) [below of=r1] {$\vdots$}
  node (r2)  [draw,rectangle,minimum width=6em,below of=v1] {}
  node (x2) [right of=r2, right] {$x_j[1..n]$}
  node (v2) [below of=r2] {$\vdots$}
  node (r3)  [draw,rectangle,minimum width=6em,below of=v2] {}
  node (x3) [right of=r3, right] {$x_m[1..n]$}
  node (u1) [draw,rectangle,right of=x1] {}
  node (v1) [right of=u1,left=-0.3em] {used$_1$}
  node (u2) [draw,rectangle,right of=x2] {}
  node (v2) [right of=u2,left=-0.3em] {used$_j$}
  node (u3) [draw,rectangle,right of=x3] {}
  node (v3) [right of=u3,left=-0.5em] {used$_m$}
  node (b1) [right of=v1, right, draw,rectangle,minimum width=6em] {}
  node (c1) [right of=b1, right] {$b_1[2..n]$}
  node (d1) [below of=b1,above] {$\ddots$}
  node (b2) [below of=d1, above, draw,rectangle,minimum width=3em,right=0em] {}
  node (c2) [right of=b2,right=-1.5em] {$b_i[(i+1)..n]$}
  node (d2) [below of=b2,above] {$\ddots$}
  node (av1) [right of= c1, right=2em,draw, rectangle] {}
  node (bv1) [right of=av1,left=0.3em] {$av_1$}
  node (vv1) [below of=av1,above] {$\vdots$}
  node (av2) at (c2-|av1) [draw, rectangle] {}
  node (bv2) [right of=av2,left=0.3em] {$av_i$}
  node (vv2) [below of=av2,above] {$\vdots$}  
  node (av3) [below of= vv2, above, draw, rectangle]  {}
  node (bv3) [right of=av3, left=0.3em] {$av_n$}  
  node (anchor) at ($(r1.west|-x1.north)+(-.1,0)$) {};
  \draw[dashed,color=gray] (anchor.center) -- node[above,color=gray] {Input} (anchor.center-|x3.south east) -- (x3.south east) -- (x3.south east-|anchor.center) -- cycle;
 \end{tikzpicture}
 \caption{Abstract memory layout. Input is the $x_j[1..n]$, all other qubits are set to 0 except $av_i$ which is set to 1. used$_j$ states whether the vector has been put in the basis. $av_i$ states whether the basis contains a vector of the form $0^{i-1}1*$. The $*$ part is stored in $b_i[(i+1)..n]$. }\label{fig:layout}
 \end{figure}
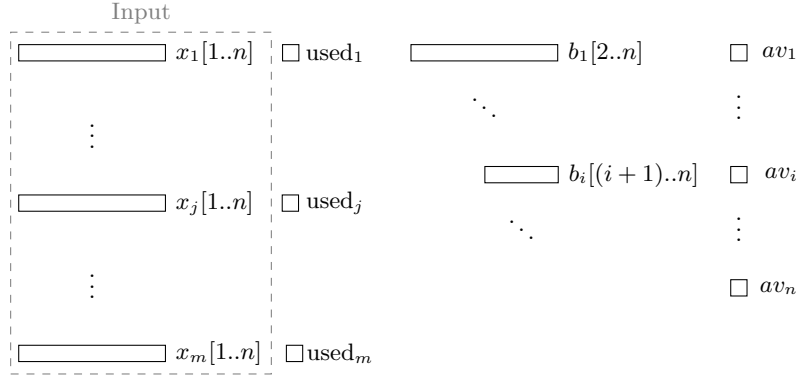

\autoref{fig:layout} represents the qubits in the algorithm.
 
\begin{algorithm}
 \begin{algorithmic}[1]
  \For{ $i$ from 1 to $n$}\label{alg:step:basis_n_loop}
  \For{ $j$ for 1 to $m$}\label{alg:step:basis_m_loop}
  \State used$_j =\text{used}_j + x_j[i] \wedge av_i$ \Comment{Do we need to insert $x_j$ to $b_i$?}\label{alg:step:basis_used_set}
  \State $av_i = av_i + x_j[i] \wedge$ used$_j$\Comment{Set $av_i$ to 0 if we insert $x_j$.}\label{alg:step:basis_av_set}
  \If{ used$_j$} \Comment{Insert $x_j$ to $b_i$.}\label{alg:step:basis_insert_x}
  \State $b_i[(i+1)..n]=b_i[(i+1)..n] + x_j[(i+1)..n]$
  \EndIf
  \If{ $x_j[i]$ } \Comment{Reduce the vector using the basis.}\label{alg:step:basis_reduce_x}
  \State{ $x_j[(i+1)..n]=x_j[(i+1)..n] + b_i[(i+1)..n]$ }
  \EndIf
  \EndFor
  \EndFor
\end{algorithmic}
\caption{Triangular basis computation}\label{alg:algebra}
\end{algorithm}

\begin{definition}
  We let $(i,j)$ denote the $j$th iteration of the inner loop in the $i$th iteration of the outer loop. We use the partial order $(i,j) \leq (k,l) \Leftrightarrow i\leq k \wedge j\leq l$, and assume that $(i,j)$ occured before $(k,l)$ if $(i,j) < (k,l)$.
\end{definition}

\begin{theorem}[Correctness of {\autoref{alg:algebra}}]
We let $\beta_i$ denote the vector $0^{i-1} \Vert \overline{av_i}\Vert b[(i+1)..n]\in \{0,1\}^n$, with the values of $av_i,b[(i+1)..n]$ at the end of \autoref{alg:algebra}.
 Then $\left\langle x_j\right\rangle = \left\langle \beta_i\right\rangle$.
\end{theorem}

\begin{proof}

 To prove the correctness of the algorithm, we begin with the following lemma:
\begin{lemma}[Algorithm invariants]
At the beginning of $(i,j)$, if $av_i = 1$, then $b_i[i+1..n] = 0^{n-i}$. If used$_j  = 1$, then $x_j[i..n] = 0^{n-i+1}$.
\end{lemma}

\begin{proof}
 We prove this by induction over $(i,j)$. We do not enforce a total order on the iterations. Here, we only need that each $(i,j)$ is computed atomically; that is, we cannot have parallel iterations with the same $i$ or $j$, and we enforce that $(i,j)$ occurs after $(k,l)$ for all $(k,l) < (i,j)$.
 
 At the beginning of $(0,0)$, $av_i = 1$ and used$_j  = 0$, hence the lemma holds.
 
 Assume that at the the beginning of $(i,j)$, the lemma holds. We now want to prove that it will still hold at the end.
 \begin{itemize}
  \item If $x_j[i] = 0$, used$_j$ and $av_i$ stay invariant. Step~\ref{alg:step:basis_insert_x} updates $b_i[(i+1)..n]$ if and only if used$_j = 1$. By the induction hypothesis, $x_j[i..n] = 0$, sp $b_i[(i+1)..n]$ is unchanged.
  \item If $x_j[i] = 1$, we must have used$_j = 0$, by the induction hypothesis.
  \begin{itemize}
   \item If $av_i = 0$, used$_j$ is not updated, hence $av_i$ is also not updated.
   \item If $av_i = 1$, then $b_i[(i+1)..n]=0$. We have used$_j$ set to 1 at Step~\ref{alg:step:basis_used_set} , $av_i$ set to 0 at Step~\ref{alg:step:basis_av_set} and $b_i[(i+1)..n]$ is set to $x_j[(i+1)..n]$ at Step~\ref{alg:step:basis_insert_x}. Step~\ref{alg:step:basis_reduce_x} reduces $x_j[(i+1)..n]$ with $b_i[(i+1)..n] = x_j[(i+1)..n]$. Hence, $x_j[(i+1)..n] = 0$, and we have that for all $k > i$, $x_j[k..n] = 0$.
  \end{itemize}
  From this, the lemma still holds after $(i,j)$.
 \end{itemize}
\end{proof}

  \begin{lemma}
   Iteration $i$ of the outer \emph{for} loop sets $\beta_i$ as the first $x_j$ with a 1 at position $i$ if any exists, and makes a partial gaussian eliminitation on all the following $x_j$ using $\beta_i$.
 \end{lemma}
 
\begin{proof}
At the beginning of iteration $i$, we must have $av_i = 1$ and $\beta_i = 0$, as these variables did not intervene earlier.

Now, while $x_j[i] = 0$, nothing happens (indeed, if used$_j = 1$, then $x_j[(i+1)..n] = 0$, by the previous lemma).

At the first $x_j[i] = 1$, we set $av_i$ to 0 and $b_i$ to $x_j[(i+1)..n]$. Hence, $\beta_i = x_j[i]$.

Then, $av_i$ and $b_i$ can no longer be modified, and we add $b[(i+1)..n]$ to $x_j[(i+1)..n]$ if $x_j[i] = 1$. This acts as a gaussian elimination on $x_j$ using $\beta_i$.\qed
\end{proof}

Hence, if we sequentially apply the previous lemma, we get one $\beta_i$ at each outer \emph{for} loop, if any such vector exists.
In the end, either the vectors are put in $b_i$ or fully reduced to 0. Hence, the theorem holds.\qed
\end{proof}

\begin{remark}[Parallel computation]\label{remark:parallel_linear_algebra}
 For the correctness of the algorithm, we only need that if $(i,j) < (k,l)$, then $(i,j)$ must be computed before $(k,l)$. This allows us to compute in parallel the steps $(i,j)$ with $i+j$ constant, as they are independent. 
\end{remark}

\subsection{Cost analysis}\label{sec:algebra_costs}

\subsubsection{Qubits.} The circuit modifies in-place its $m\times n$ qubit input, though it needs $m+n(n+1)/2$ auxiliary qubits for $b$, used, and $av$. We also use another $n(n-1)$ auxiliary qubits to reduce the depth of row reductions, as detailed below.

\subsubsection{Gate count.}Steps~\ref{alg:step:basis_used_set} and~\ref{alg:step:basis_av_set} require just one Toffoli gate and are repeated $mn$ times. Inserting $x_j$ at Step~\ref{alg:step:basis_insert_x} requires $n-i$ Toffoli gates, as does Step~\ref{alg:step:basis_reduce_x}. Summed over all $i$, and repeated $m$ times, gives a total of $mn^2+mn$ Toffoli gates to compute the triangular basis.

\subsubsection{Depth.} As Remark~\ref{remark:parallel_linear_algebra} indicates, we can compute two iterations $(i,j)$ and $(i',j')$ in parallel if $i+j=i'+j'$. Hence, we only need to perform $m+n$ iterations sequentially.

Iteration $(i,j)$ has a naive depth of $2(n-i+1)+2$, as inserting and reducing $x_j$ are controlled by single qubits, so we must apply each Toffoli  sequentially. However, we can fan out the control to apply the Toffolis simultaneously. This means a depth of $\ceiling{\log_2(n-i+1)}+4$, though this is what requires the extra $n(n-1)$ auxiliary qubits.

When reducing $x_j$, once we have modified $x_j[i+1]$, we can begin the next iteration with $(i+1,j)$, and reduce $x_j[(i+2)\dots n]$ simultaneously. However, the same logic does not apply to inserting $x_j$ into the basis; we need to finish with used$_j$ before the next iteration modifies it. 

This gives us a total circuit depth of $O((m+n)\lg(n))$. The specific constants will depend on our cost model, the structure of the fanout, and the choice of Toffoli gate. We used linear regression on the results from Q\# to estimate the concrete asymptotics.

\subsection{Final steps}

\subsubsection{Rank computation.}
Once we have the triangular basis, we only need to check if the basis has a full rank, which only requires testing whether all $av_i$ bits are set to 0. 

\subsubsection{Computing orthogonal vectors.}
While this is not directly useful here, given the triangular basis we could easily compute a vector orthogonal to it, at a cost of $n$ CNOT and 
$n^2-n$ Toffoli. The idea is to choose the bit $i$, beginning with the last bit, such that the vector we compute is orthogonal to the basis vectors $i$ to $n$. As the basis is in triangular form, we can sequentially compute the vector. The only freedom we have is on the values we put when the vector $i$ is missing in the basis. If we only need one vector, we can simply put 1 in that case. This is \autoref{alg:dual}.

\begin{algorithm}
 \begin{algorithmic}[1]
  \For{ $i$ from $n$ to 1}
   \State $out[i] = av_i$ \Comment{Put a 1 if basis empty}
   \For{ $j$ from $i+1$ to $n$}
    \State $out[i]=out[i] + out[j] \wedge b_i[j]$\Comment{Ensure orthogonality}
  \EndFor
  \EndFor
\end{algorithmic}
\caption{Orthogonal vector computation}\label{alg:dual}
\end{algorithm}
This needs more work to compute a basis of the dual in a larger dimension, as the pattern of values we choose must form a free family.

\subsubsection{Solving linear equations.}
The same approach can solve general boolean systems of linear equations: instead of the equation $\sum_{i=1}^n a_i b_i = \epsilon$, we can consider $\sum_{i=1}^n a_i b_i + \epsilon b_{n+1} = 0$, and force the final solution to have $b_{n+1} = 1$. If we only need to know if the system is solvable, then we only need to check if $av_{n+1} = 1$, as if it is equal to 0, any solution of the equation system must fulfill $b_{n+1} = 0$.

\section{Reversible implementations of quantum primitives}\label{sec:circuits}
\subsection{Design Philosophy}

To apply our attack, we implement an operator with the following general shape:
\[
 \ket{x}\ket{i}\ket{E(x)} \mapsto \ket{x}\ket{i}\ket{E(x)\oplus f(i,x)}\enspace.
\]

Thus, there is little reason for us to prefer an in-place encryption algorithm, since we need to preserve the input for proper interference in Simon's algorithm. However, the permutations we consider are all iterated designs containing multiple rounds of some simpler permutation. If a single round is out-of-place, we either need to double our computational cost to uncompute as we proceed, or allocate fresh qubits for every round; hence, we tried to find in-place circuits.

Some permutations use small S-boxes of 4 to 5 bits. We could use a table look-up, but this is out-of-place and has cost linear in the table size (e.g., 16 AND operations for 4 bits). Instead we found optimized in-place circuits, inspired by masked implementations of block ciphers, which also use a model in which XOR is cheap and AND is expensive.

In depth-limited Grover-like algorithms, the most efficient oracle design makes strong trade-offs of depth against width. However, the Q\# resource estimator will not reuse qubits when optimizing for depth. That is, if each permutation round needed to borrow and release 10 qubits, and a cipher ran for 80 rounds, Q\# would count 800 extra qubits. To avoid this issue, we used a width-optimizing compiler, which always prefers to reuse qubits, even if that means delaying other operations. Thanks to our in-place implementations, neither issue has a large effect on our results.

\subsection{Simon-specific optimizations}\label{sec:simon_optimizations}
The primitive circuits we implement have some relaxed constraints, which allows us to compute slightly different (and cheaper) functions.

\subsubsection{Shorter output.} From \autoref{thm:off-rand}, we can afford to have a short output, which will be in practice of 11 bits. This allows us to not compute some of the output bits, and in general we can at least avoid the computation of most of the final non-linear layer.

\subsubsection{Linear combination.} For our attacks, we have the general property
\[
 f(i,x) = E(x\oplus s) \oplus c \enspace.
\]
 We can remark that for any affine function $\phi$, $\phi\circ f$ and $\phi\circ E$ will have the same general property:
 \[
  \phi\circ f(i,x) = \phi\circ E(x\oplus s) \oplus c'
 \]

 Hence, we can apply any affine function to the output of our function (as long as its output is long enough). This actually generalizes the previous property, as truncation is linear.
 
 Overall, we can remove many operations in the last rounds: the ones that either do not influence the bits we're interested in, or only act linearly on them.
 
 \subsubsection{Partially fixed input.} 
\begin{figure}
\begin{subfigure}[b]{0.45\textwidth}
\centering
 \begin{tikzpicture}
  \node (y) at (0,0) {$y$};
  \node[below of=y,above=0.5em] (x) {$x$};
  \path (x) -- node (middle) {} (y);
  \node [block,right = of middle, minimum height=4em] (f){$f$};
  \node[right=of f] (out) {$f(k,x||y)$};
  \node[above of=f, above] (k) {$k$};
  \draw (y) -- (f.west|-y);
  \draw (x) -- (f.west|-x);
  \draw (k) -- (f);
  \draw (f) -- (out);
 \end{tikzpicture}
\caption{General shape of the functions we implement}\label{fig:atk-shape}
\end{subfigure}
\def\fcolor{gray}
\def\gcolor{blue}
\begin{subfigure}[b]{0.45\textwidth}
\centering
 \begin{tikzpicture}
  \node[color=\gcolor] (y) at (0,0) {$y$};
  \node[below of=y,above] (x) {$x$};
  \node[block, right of=y,color=\gcolor] (g) {$g$};
  \node[right of=x,minimum size = 2em] (d) {};
  \node (m) at ($(g.south east)+(0.2,-0.2)$) {};
  \node[right of = d,minimum size = 2em] (f) {};
  \node[right of = g,minimum size = 2em] (h) {};
  \path (f) -- node (fm) {$f'$~~} (h);
  \path (g) -- node (kb) {} (h);
  \node[above of=kb] {\raisebox{-1.3em}{$k$}};
  \draw[color=\gcolor] (k) -| (g);
  \draw (k) -| (h); 
  \node[right=of fm] (out) {$f(k,x||y)$};
  \draw (m.center) -- (m.center|-g.north) -- (f.east|-g.north) -- (f.south east) -- (f.south-|g.west) -- (m.center-|g.west) -- (m.center);
  \draw (x) -- (d);
  \draw[color=\gcolor] (y) -- (g) -- (m.center|-g);
  \draw (out) -- (out-|h.east);
  \draw[dashed,color=\fcolor] ($(g.north west)+(-0.1,0.1)$) -- ($(h.north east)+(0.1,0.1)$) --($(f.south east)+(0.1,-0.1)$) -- node[below] {$f$} ($(d.south west)+(-0.1,-0.1)$) -- cycle ;
 \end{tikzpicture}
\caption{Structure suitable for optimization\newline}\label{fig:opt-shape}
\end{subfigure}\caption{Functions we use in Simon's algorithm.}
\end{figure}
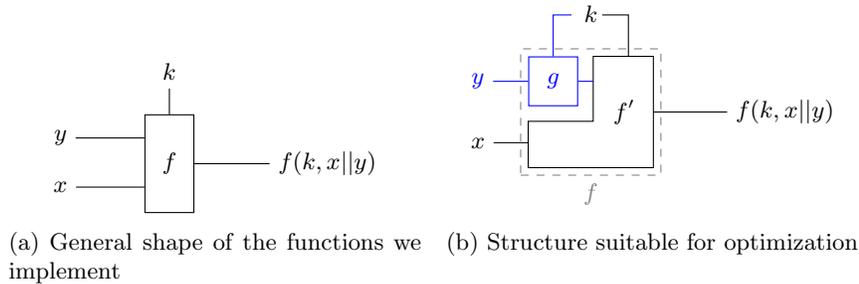

We can split the variable $i$ on which we do a quantum search into two: $y$, which corresponds to the part of the message which is fixed, and $k$, which is a secret we must guess completely. For Even-Mansour, $k$ is empty, and for the FX construction, $y$ can be empty. The general shape is presented on \autoref{fig:atk-shape}.
 
Moreover, the design of the function transforms the input in-place and bijectively. This means we can decompose the full function $f$ into $f(k,x,y) = f'(k,x,g(k,y))$, as in~\autoref{fig:opt-shape}. With this specific structure, the output of $g$ will be identical for all the parallel computations of $f$. As $y$ is guessed by the quantum search, we can afford to only compute $g$ once for all the parallel computations of $f$. This saves us some computation, depending on how fast the input bits diffuse. We found ways to save part of the first linear layer and a few S-boxes.

We go further and remark that in many cases, the mapping $y\mapsto g(k,y)$ will be a permutation. Hence, instead of applying the quantum search to $f$ to find $k$ and $y$, we search $f'$ to find $k$ and $g(k,y)$. Once we find $g(k,y)$ and $k$, it is easy to invert and find $y$. This allows us to completely remove all the operations that only operate on the bits of $y$ from the quantum circuit.

\subsubsection{Summary.} We can leverage the specific structure of the problem to reduce the computational cost of $f$. These optimizations rely on the limits of the diffusion in some iterated constructions. In practice, for the constructions we considered, they save a cost equivalent to 1 to 2 rounds, which becomes completely negligible for constructions with a very large number of rounds. Nevertheless, these optimizations are independent of the actual implementation of the quantum circuit, and can always be applied.

\subsection{Chaskey}\label{sec:chaskey-circuit}

The \Chaskey{} permutation has an ARX structure: it uses only XOR, bit rotation, and modular addition. All of these can be implemented in-place on a quantum computer, and efficient circuits for them are already available~\cite{PQCRYPTO:HJNRS20}. We use the adder with the fewest T operations~\cite{Q:Gidney18}. The quantum circuit for the permutation is practically identical to the classical circuit.

Optimizations from Section~\ref{sec:simon_optimizations} for a shorter output are particularly effective, detailed in \autoref{fig:ciphers:chaskey_first} and~\ref{fig:ciphers:chaskey}. We save a fourth of the operation in the first round thanks to the partially fixed input, shown in~\autoref{fig:ciphers:chaskey_first}.
\autoref{fig:ciphers:chaskey} presents the last two rounds of the truncated permutation. Once it is computed, we copy out bits from 5 to 15 \emph{and} from 37 to 47 into the output register before uncomputing. This has the same effect as the CNOT highlighted in green in \autoref{fig:ciphers:chaskey}, but saves uncomputation. The total effect is 18\% in depth  and operation savings for 8 rounds and 12.5\% for 12 rounds.

\begin{circuit}[h]
\centering
\resizebox{!}{0.18\textwidth}{
\input{qpic/chaskey-first.tikz}
}
\caption{The \Chaskey{} permutation round. Operations in {\color{red}red} can be removed in the first round.}\label{fig:ciphers:chaskey_first}
\end{circuit}

\begin{circuit}[h]
\centering
\resizebox{!}{0.18\textwidth}{
\input{qpic/chaskey-unoptimized.tikz}
}
\caption{The last two rounds of \Chaskey{}'s permutation. Operations in {\color{red}red} can be removed; those in {\color{blue}blue} can be inverted with a linear operation applied to the known ciphertexts; the {\color{green!50!black}green} operation can be done only when copying out; the additions \colorbox{red!30!blue!30!white}{highlighted in purple} and the purple CNOT only need the least significant 16 bits.}\label{fig:ciphers:chaskey}
\end{circuit}

\subsection{Prince}

Internally, \Prince{} uses a keyed permutation of 12 rounds, where each round XORs round constants, applies an S-box to each nibble, multiplies the state by a binary matrix, and XORs the key (\autoref{fig:prince_circuit}). 

We implemented \Prince{} in-place with the S-box decomposition from~\cite{EPRINT:BoiKneNik18}, which only requires 6 Toffoli operations per S-box (\autoref{fig:prince_sbox}).

We perform a PLU decomposition for the linear layer as well as the affine layers in the S-box decomposition, as in~\cite{EC:JNRV20}. 

Round 9 only needs to apply the S-box to nibbles 3, 6, 9, and 12. Then in round 10, we only need to use those bits of the key and the round constant. We only apply the part of the linear layer necessary to compute these nibbles, and then the row shift puts these nibbles in the first 16 bits. We finish with an S-box on these bits. This saves us 13.5\% of all operations, though provides negligible depth reduction.

\begin{circuit}[h]
\centering
\resizebox{!}{0.12\textwidth}{
\input{qpic/prince_permutation.tikz}
}
\caption{\Prince's permutation, where $S$ is the S-box, $M$ is multiplication by a fixed binary matrix $M'$, and $RC_i$ are round constants.}\label{fig:prince_circuit}
\end{circuit}

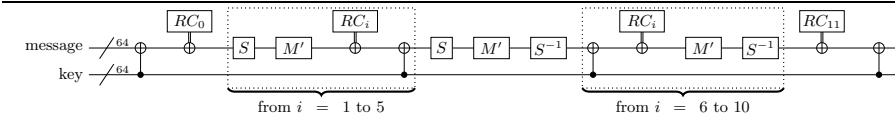
\begin{circuit}[h]
\centering
\resizebox{!}{0.13\textwidth}{
\input{qpic/prince_sbox.tikz}
}
\caption{\Prince's S-box, applied to 4 qubits.}\label{fig:prince_sbox}
\end{circuit}

\subsection{Elephant-160/176}

\Elephant-160 and 176 use the \spongent{} permutation~\cite{CHES:BKLTVV11}, with respectively 80 and 96 rounds (\autoref{fig:elephant_circuit}).

The first step of each round is an XOR with a fixed sequence of strings $C_i$, which requires only a series of $X$ operations. The next step is an S-box layer. We implemented it in-place using a masking-friendly decomposition that only required 4 Toffoli operations (\autoref{fig:spongent_sbox}), using the fact that 4 bit S-boxes are fully classified and their decomposition as a composition of quadratic functions is known~\cite{PhD:DeCanniere,CHES:BNNRS12,TI-tools}. The final step is a permutation, which can be done by the classical computer with no extra quantum operations. 

Input and output optimizations are less effective here because \Elephant{} repeats so many rounds. We still limit the final layer of the S-box to only the bits we use in the output, resulting in 1.8\% and 1.7\% operation savings for Elephant-160 and 176, respectively, with no depth improvement.

\begin{circuit}[h]
\resizebox{!}{0.2\textwidth}{
\providecommand{\ket}[1]{\left\vert #1\right\rangle}
\begin{tikzpicture}[scale=1.25000,x=1pt,y=1pt]
\draw(65, 45) node[above, text centered] {\Large{\spongent}};
\draw[color=black] (-25.000000, -30.00000) rectangle (155.000000, 45.00000);
\draw[color=black] (0.000000,15.000000) -- (152.000000,15.000000);
\draw[color=black] (0.000000,15.000000) node[left] {$\ket{m}$};
\draw (9.500000, 9.000000) -- (17.500000, 21.000000);
\draw (15.500000, 18.000000) node[right] {$\scriptstyle{n}$};
\draw (65.000000,30.000000) -- (65.000000,15.000000);
\draw (66.000000,30.000000) -- (66.000000,15.000000);
\begin{scope}
\draw[fill=white] (65.500000, 30.000000) +(-45.000000:45.961941pt and 8.485281pt) -- +(45.000000:45.961941pt and 8.485281pt) -- +(135.000000:45.961941pt and 8.485281pt) -- +(225.000000:45.961941pt and 8.485281pt) -- cycle;
\clip (65.500000, 30.000000) +(-45.000000:45.961941pt and 8.485281pt) -- +(45.000000:45.961941pt and 8.485281pt) -- +(135.000000:45.961941pt and 8.485281pt) -- +(225.000000:45.961941pt and 8.485281pt) -- cycle;
\draw (65.500000, 30.000000) node {$\scriptstyle rev(C_i)\Vert 0^{n-13}\Vert C_{i}$};
\end{scope}
\begin{scope}
\draw[fill=white] (65.500000, 15.000000) circle(3.000000pt);
\clip (65.500000, 15.000000) circle(3.000000pt);
\draw (62.500000, 15.000000) -- (68.500000, 15.000000);
\draw (65.500000, 12.000000) -- (65.500000, 18.000000);
\end{scope}
\begin{scope}
\draw[fill=white] (116.000000, 15.000000) +(-45.000000:8.485281pt and 8.485281pt) -- +(45.000000:8.485281pt and 8.485281pt) -- +(135.000000:8.485281pt and 8.485281pt) -- +(225.000000:8.485281pt and 8.485281pt) -- cycle;
\clip (116.000000, 15.000000) +(-45.000000:8.485281pt and 8.485281pt) -- +(45.000000:8.485281pt and 8.485281pt) -- +(135.000000:8.485281pt and 8.485281pt) -- +(225.000000:8.485281pt and 8.485281pt) -- cycle;
\draw (116.000000, 15.000000) node {$S$};
\end{scope}
\begin{scope}
\draw[fill=white] (140.000000, 15.000000) +(-45.000000:8.485281pt and 28.284271pt) -- +(45.000000:8.485281pt and 28.284271pt) -- +(135.000000:8.485281pt and 28.284271pt) -- +(225.000000:8.485281pt and 28.284271pt) -- cycle;
\clip (140.000000, 15.000000) +(-45.000000:8.485281pt and 28.284271pt) -- +(45.000000:8.485281pt and 28.284271pt) -- +(135.000000:8.485281pt and 28.284271pt) -- +(225.000000:8.485281pt and 28.284271pt) -- cycle;
\draw (140.000000, 15.000000) node {$\rotatebox{90}{PLayer}$};
\end{scope}
\draw[draw opacity=1.000000,fill opacity=0.200000,color=black,dotted] (30.000000,37.500000) rectangle (149.000000,-7.500000);
\draw[draw opacity=1.000000,fill opacity=0.200000,color=black,dotted] (30.000000,37.500000) rectangle (149.000000,-7.500000);
\draw[decorate,decoration={brace,mirror,amplitude = 4.000000pt},very thick] (30.000000,-7.500000) -- (149.000000,-7.500000);
\draw (89.500000, -11.500000) node[text width=144pt,below,text centered] {from $i=1$ to $\{80,96\}$};
\end{tikzpicture}
}\resizebox{!}{0.2\textwidth}{
\providecommand{\ket}[1]{\left\vert #1\right\rangle}
\begin{tikzpicture}[scale=1.000000,x=1pt,y=1pt]
\draw(92.5,45) node[above, text centered] {\Large{\keccak}};
\draw[color=black] (-30.000000, -30.00000) rectangle (185.000000, 45.00000);
\draw[color=black,rounded corners=4.000000pt] (46.000000,30.000000) -- (122.000000,30.000000) -- (129.500000,22.500000);
\draw[color=black,rounded corners=4.000000pt] (129.500000,22.500000) -- (137.000000,15.000000) -- (175.000000,15.000000);
\draw[color=black,rounded corners=4.000000pt] (0.000000,15.000000) -- (122.000000,15.000000) -- (129.500000,22.500000);
\draw[color=black,rounded corners=4.000000pt] (129.500000,22.500000) -- (137.000000,30.000000) -- (156.500000,30.000000);
\draw[color=black] (0.000000,15.000000) node[left] {$\ket{m}$};
\draw (3.000000, 9.000000) -- (11.000000, 21.000000);
\draw (9.000000, 18.000000) node[right] {$\scriptstyle{200}$};
\begin{scope}
\draw[fill=white] (46.000000, 15.000000) +(-45.000000:28.284271pt and 8.485281pt) -- +(45.000000:28.284271pt and 8.485281pt) -- +(135.000000:28.284271pt and 8.485281pt) -- +(225.000000:28.284271pt and 8.485281pt) -- cycle;
\clip (46.000000, 15.000000) +(-45.000000:28.284271pt and 8.485281pt) -- +(45.000000:28.284271pt and 8.485281pt) -- +(135.000000:28.284271pt and 8.485281pt) -- +(225.000000:28.284271pt and 8.485281pt) -- cycle;
\draw (46.000000, 15.000000) node {$\pi\circ\rho\circ\theta$};
\end{scope}
\draw[color=black] (53.500000,30.000000) node[fill=white,left,minimum height=15.000000pt,minimum width=15.000000pt,inner sep=0pt] {\phantom{$\ket{0}$}};
\draw[color=black] (53.500000,30.000000) node[left] {$\ket{0}$};
\draw (73.000000, 24.000000) -- (81.000000, 36.000000);
\draw (79.000000, 33.000000) node[right] {$\scriptstyle{200}$};
\draw (104.000000,30.000000) -- (104.000000,15.000000);
\begin{scope}
\draw[fill=white] (104.000000, 22.500000) +(-45.000000:8.485281pt and 19.091883pt) -- +(45.000000:8.485281pt and 19.091883pt) -- +(135.000000:8.485281pt and 19.091883pt) -- +(225.000000:8.485281pt and 19.091883pt) -- cycle;
\clip (104.000000, 22.500000) +(-45.000000:8.485281pt and 19.091883pt) -- +(45.000000:8.485281pt and 19.091883pt) -- +(135.000000:8.485281pt and 19.091883pt) -- +(225.000000:8.485281pt and 19.091883pt) -- cycle;
\draw (104.000000, 22.500000) node {$\chi$};
\end{scope}
\draw[color=black] (149.000000,30.000000) node[fill=white,right,minimum height=15.000000pt,minimum width=15.000000pt,inner sep=0pt] {\phantom{$\ket{0}$}};
\draw[color=black] (149.000000,30.000000) node[right] {$\ket{0}$};
\draw (149.000000,15.000000) -- (149.000000,0.000000);
\draw (150.000000,15.000000) -- (150.000000,0.000000);
\begin{scope}
\draw[fill=white] (149.500000, -0.000000) +(-45.000000:17.677670pt and 8.485281pt) -- +(45.000000:17.677670pt and 8.485281pt) -- +(135.000000:17.677670pt and 8.485281pt) -- +(225.000000:17.677670pt and 8.485281pt) -- cycle;
\clip (149.500000, -0.000000) +(-45.000000:17.677670pt and 8.485281pt) -- +(45.000000:17.677670pt and 8.485281pt) -- +(135.000000:17.677670pt and 8.485281pt) -- +(225.000000:17.677670pt and 8.485281pt) -- cycle;
\draw (149.500000, -0.000000) node {$\iota_i$};
\end{scope}
\begin{scope}
\draw[fill=white] (149.500000, 15.000000) circle(3.000000pt);
\clip (149.500000, 15.000000) circle(3.000000pt);
\draw (146.500000, 15.000000) -- (152.500000, 15.000000);
\draw (149.500000, 12.000000) -- (149.500000, 18.000000);
\end{scope}
\draw[draw opacity=1.000000,fill opacity=0.200000,color=black,dotted] (23.000000,37.500000) rectangle (165.000000,-7.500000);
\draw[draw opacity=1.000000,fill opacity=0.200000,color=black,dotted] (23.000000,37.500000) rectangle (165.000000,-7.500000);
\draw[decorate,decoration={brace,mirror,amplitude = 4.000000pt},very thick] (23.000000,-7.500000) -- (165.000000,-7.500000);
\draw (94.00000, -11.500000) node[text width=144pt,below,text centered] {from $i=1$ to $18$};
\end{tikzpicture}
}
\caption{\Elephant{}'s permutations.}\label{fig:elephant_circuit}
\end{circuit}

\begin{circuit}[h]
\centering
\resizebox{!}{0.15\textwidth}{
\providecommand{\ket}[1]{\left\vert #1\right\rangle}
\begin{tikzpicture}[scale=1.000000,x=1pt,y=1pt]
\filldraw[color=white] (0.000000, -7.500000) rectangle (303.000000, 52.500000);
\draw[color=black] (0.000000,45.000000) -- (60.000000,45.000000);
\draw[color=black] (60.000000,45.000000) -- (67.500000,45.000000);
\draw[color=black] (67.500000,45.000000) -- (75.000000,45.000000);
\draw[color=black,rounded corners=4.000000pt] (75.000000,45.000000) -- (141.000000,45.000000) -- (148.500000,22.500000);
\draw[color=black,rounded corners=4.000000pt] (148.500000,22.500000) -- (156.000000,0.000000) -- (264.000000,0.000000) -- (271.500000,15.000000);
\draw[color=black,rounded corners=4.000000pt] (271.500000,15.000000) -- (279.000000,30.000000) -- (303.000000,30.000000);
\draw[color=black] (0.000000,45.000000) node[left] {$\ket{m_0}$};
\draw[color=black] (0.000000,30.000000) -- (60.000000,30.000000);
\draw[color=black] (60.000000,30.000000) -- (67.500000,30.000000);
\draw[color=black] (67.500000,30.000000) -- (75.000000,30.000000);
\draw[color=black] (75.000000,30.000000) -- (141.000000,30.000000);
\draw[color=black] (141.000000,30.000000) -- (148.500000,30.000000);
\draw[color=black] (148.500000,30.000000) -- (156.000000,30.000000);
\draw[color=black,rounded corners=4.000000pt] (156.000000,30.000000) -- (264.000000,30.000000) -- (271.500000,22.500000);
\draw[color=black,rounded corners=4.000000pt] (271.500000,22.500000) -- (279.000000,15.000000) -- (303.000000,15.000000);
\draw[color=black] (0.000000,30.000000) node[left] {$\ket{m_1}$};
\draw[color=black,rounded corners=4.000000pt] (0.000000,15.000000) -- (60.000000,15.000000) -- (67.500000,7.500000);
\draw[color=black,rounded corners=4.000000pt] (67.500000,7.500000) -- (75.000000,0.000000) -- (141.000000,0.000000) -- (148.500000,22.500000);
\draw[color=black,rounded corners=4.000000pt] (148.500000,22.500000) -- (156.000000,45.000000) -- (264.000000,45.000000);
\draw[color=black] (264.000000,45.000000) -- (271.500000,45.000000);
\draw[color=black] (271.500000,45.000000) -- (279.000000,45.000000);
\draw[color=black] (279.000000,45.000000) -- (303.000000,45.000000);
\draw[color=black] (0.000000,15.000000) node[left] {$\ket{m_2}$};
\draw[color=black,rounded corners=4.000000pt] (0.000000,0.000000) -- (60.000000,0.000000) -- (67.500000,7.500000);
\draw[color=black,rounded corners=4.000000pt] (67.500000,7.500000) -- (75.000000,15.000000) -- (141.000000,15.000000);
\draw[color=black] (141.000000,15.000000) -- (148.500000,15.000000);
\draw[color=black] (148.500000,15.000000) -- (156.000000,15.000000);
\draw[color=black,rounded corners=4.000000pt] (156.000000,15.000000) -- (264.000000,15.000000) -- (271.500000,7.500000);
\draw[color=black,rounded corners=4.000000pt] (271.500000,7.500000) -- (279.000000,0.000000) -- (303.000000,0.000000);
\draw[color=black] (0.000000,0.000000) node[left] {$\ket{m_3}$};
\draw (9.000000,45.000000) -- (9.000000,15.000000);
\filldraw (9.000000, 15.000000) circle(1.500000pt);
\begin{scope}
\draw[fill=white] (9.000000, 45.000000) circle(3.000000pt);
\clip (9.000000, 45.000000) circle(3.000000pt);
\draw (6.000000, 45.000000) -- (12.000000, 45.000000);
\draw (9.000000, 42.000000) -- (9.000000, 48.000000);
\end{scope}
\draw (27.000000,30.000000) -- (27.000000,15.000000);
\filldraw (27.000000, 30.000000) circle(1.500000pt);
\begin{scope}
\draw[fill=white] (27.000000, 15.000000) circle(3.000000pt);
\clip (27.000000, 15.000000) circle(3.000000pt);
\draw (24.000000, 15.000000) -- (30.000000, 15.000000);
\draw (27.000000, 12.000000) -- (27.000000, 18.000000);
\end{scope}
\begin{scope}
\draw[fill=white] (27.000000, 45.000000) circle(3.000000pt);
\clip (27.000000, 45.000000) circle(3.000000pt);
\draw (24.000000, 45.000000) -- (30.000000, 45.000000);
\draw (27.000000, 42.000000) -- (27.000000, 48.000000);
\end{scope}
\draw (45.000000,30.000000) -- (45.000000,0.000000);
\filldraw (45.000000, 0.000000) circle(1.500000pt);
\begin{scope}
\draw[fill=white] (45.000000, 30.000000) circle(3.000000pt);
\clip (45.000000, 30.000000) circle(3.000000pt);
\draw (42.000000, 30.000000) -- (48.000000, 30.000000);
\draw (45.000000, 27.000000) -- (45.000000, 33.000000);
\end{scope}
\draw (90.000000,45.000000) -- (90.000000,0.000000);
\filldraw (90.000000, 0.000000) circle(1.500000pt);
\filldraw (90.000000, 30.000000) circle(1.500000pt);
\begin{scope}
\draw[fill=white] (90.000000, 45.000000) circle(3.000000pt);
\clip (90.000000, 45.000000) circle(3.000000pt);
\draw (87.000000, 45.000000) -- (93.000000, 45.000000);
\draw (90.000000, 42.000000) -- (90.000000, 48.000000);
\end{scope}
\draw (108.000000,30.000000) -- (108.000000,0.000000);
\filldraw (108.000000, 0.000000) circle(1.500000pt);
\filldraw (108.000000, 15.000000) circle(1.500000pt);
\begin{scope}
\draw[fill=white] (108.000000, 30.000000) circle(3.000000pt);
\clip (108.000000, 30.000000) circle(3.000000pt);
\draw (105.000000, 30.000000) -- (111.000000, 30.000000);
\draw (108.000000, 27.000000) -- (108.000000, 33.000000);
\end{scope}
\draw (126.000000,30.000000) -- (126.000000,0.000000);
\filldraw (126.000000, 30.000000) circle(1.500000pt);
\begin{scope}
\draw[fill=white] (126.000000, 0.000000) circle(3.000000pt);
\clip (126.000000, 0.000000) circle(3.000000pt);
\draw (123.000000, 0.000000) -- (129.000000, 0.000000);
\draw (126.000000, -3.000000) -- (126.000000, 3.000000);
\end{scope}
\draw (171.000000,30.000000) -- (171.000000,0.000000);
\filldraw (171.000000, 0.000000) circle(1.500000pt);
\filldraw (171.000000, 30.000000) circle(1.500000pt);
\begin{scope}
\draw[fill=white] (171.000000, 15.000000) circle(3.000000pt);
\clip (171.000000, 15.000000) circle(3.000000pt);
\draw (168.000000, 15.000000) -- (174.000000, 15.000000);
\draw (171.000000, 12.000000) -- (171.000000, 18.000000);
\end{scope}
\draw (189.000000,30.000000) -- (189.000000,0.000000);
\filldraw (189.000000, 0.000000) circle(1.500000pt);
\filldraw (189.000000, 15.000000) circle(1.500000pt);
\begin{scope}
\draw[fill=white] (189.000000, 30.000000) circle(3.000000pt);
\clip (189.000000, 30.000000) circle(3.000000pt);
\draw (186.000000, 30.000000) -- (192.000000, 30.000000);
\draw (189.000000, 27.000000) -- (189.000000, 33.000000);
\end{scope}
\draw (207.000000,45.000000) -- (207.000000,15.000000);
\filldraw (207.000000, 45.000000) circle(1.500000pt);
\begin{scope}
\draw[fill=white] (207.000000, 15.000000) circle(3.000000pt);
\clip (207.000000, 15.000000) circle(3.000000pt);
\draw (204.000000, 15.000000) -- (210.000000, 15.000000);
\draw (207.000000, 12.000000) -- (207.000000, 18.000000);
\end{scope}
\draw (225.000000,45.000000) -- (225.000000,0.000000);
\filldraw (225.000000, 0.000000) circle(1.500000pt);
\begin{scope}
\draw[fill=white] (225.000000, 45.000000) circle(3.000000pt);
\clip (225.000000, 45.000000) circle(3.000000pt);
\draw (222.000000, 45.000000) -- (228.000000, 45.000000);
\draw (225.000000, 42.000000) -- (225.000000, 48.000000);
\end{scope}
\draw (231.000000,30.000000) -- (231.000000,15.000000);
\filldraw (231.000000, 15.000000) circle(1.500000pt);
\begin{scope}
\draw[fill=white] (231.000000, 30.000000) circle(3.000000pt);
\clip (231.000000, 30.000000) circle(3.000000pt);
\draw (228.000000, 30.000000) -- (234.000000, 30.000000);
\draw (231.000000, 27.000000) -- (231.000000, 33.000000);
\end{scope}
\draw (249.000000,30.000000) -- (249.000000,0.000000);
\filldraw (249.000000, 30.000000) circle(1.500000pt);
\begin{scope}
\draw[fill=white] (249.000000, 0.000000) circle(3.000000pt);
\clip (249.000000, 0.000000) circle(3.000000pt);
\draw (246.000000, 0.000000) -- (252.000000, 0.000000);
\draw (249.000000, -3.000000) -- (249.000000, 3.000000);
\end{scope}
\begin{scope}
\draw[fill=white] (294.000000, 45.000000) circle(3.000000pt);
\clip (294.000000, 45.000000) circle(3.000000pt);
\draw (291.000000, 45.000000) -- (297.000000, 45.000000);
\draw (294.000000, 42.000000) -- (294.000000, 48.000000);
\end{scope}
\begin{scope}
\draw[fill=white] (294.000000, 15.000000) circle(3.000000pt);
\clip (294.000000, 15.000000) circle(3.000000pt);
\draw (291.000000, 15.000000) -- (297.000000, 15.000000);
\draw (294.000000, 12.000000) -- (294.000000, 18.000000);
\end{scope}
\begin{scope}
\draw[fill=white] (294.000000, 0.000000) circle(3.000000pt);
\clip (294.000000, 0.000000) circle(3.000000pt);
\draw (291.000000, 0.000000) -- (297.000000, 0.000000);
\draw (294.000000, -3.000000) -- (294.000000, 3.000000);
\end{scope}
\end{tikzpicture}
}
\caption{The \spongent{} S-box.}\label{fig:spongent_sbox}
\end{circuit}

\begin{circuit}[h]
\centering
\resizebox{!}{0.3\textwidth}{
\input{qpic/keccak_chi.tikz}
}
\caption{\keccak's $\chi$ function.}\label{fig:keccak_chi}
\end{circuit}

\subsection{Elephant-200}
\Elephant-200 uses a \keccak{} permutation, with a block length of 200. Each \keccak{} round starts with 3 linear functions, $\theta$, $\rho$, and $\pi$. We used a PLU decomposition of all three functions to perform them in-place. After these is the non-linear function $\chi$. We adapt the circuit from the \keccak{} implementation; however, it is out-of-place, so we also adapted a circuit for $\chi^{-1}$ from~\cite{GITHUB:HofAssKel17} (\autoref{fig:keccak_chi}). We apply the adjoint of this circuit to uncompute the input to $\chi$, then release these qubits. Since $\chi^{-1}$ is mostly AND operations, their adjoint can be done cheaply using measurements~\cite{PRA:Jones13,Q:Gidney18}. The final function is $\iota$, which simply XORs a constant onto the state, which requires only $X$ operations.

Here we can also limit the non-linear $\chi$ in the last round, for 5\% T-operation savings and 1.6\% savings over all operations.

\begin{table}
\setlength{\tabcolsep}{2pt}
\begin{tabular}{c c c c c c c c c }
\toprule
\multirow{2}{*}{Cipher}& \multirow{2}{*}{\begin{tabular}{c}Block\\Size\end{tabular}} &\multicolumn{4}{ c }{Operations} & \multicolumn{2}{c}{Depth} &\multirow{2}{*}{Qubits}\\
\cmidrule(lr){3-6}\cmidrule(lr){7-8}
 & & CNOT & 1QC & T & M & T & All & \\
\midrule
\Chaskey{}{}-8 & 128 & $1.81\cdot 2^{14}$ & 	$1.14\cdot 2^{13}$ & 	$1.63\cdot 2^{12}$ & $1.75\cdot 2^{10}$ & $1.68\cdot 2^{10}$ & $1.37\cdot 2^{14}$ & $160$\\
\Chaskey{}{}-12 & 128 & $1.46\cdot 2^{15}$ & 	$1.82\cdot 2^{13}$ & 	$1.31\cdot 2^{13}$ & $1.38\cdot 2^{11}$ & $1.36\cdot 2^{11}$ & $1.11\cdot 2^{15}$ & $160$\\
\midrule
\Prince & 64 & $1.22\cdot 2^{15}$ & 	$1.60\cdot 2^{12}$ & 	$1.68\cdot 2^{13}$ & 		$0$ & $1.64\cdot 2^{11}$ & $1.09\cdot 2^{7}$ & 128 \\
\midrule
\multirow{3}{*}{\Elephant{}} & 160 & $1.71\cdot 2^{18}$ & 	$1.17\cdot 2^{16}$ & 	$1.34\cdot 2^{17}$ & 		$0$ & $1.56\cdot 2^{11}$ & $1.29\cdot 2^{14}$ & 160 \\
 & 176 & $1.05\cdot 2^{19}$ & 	$1.45\cdot 2^{16}$ & 	$1.66\cdot 2^{17}$ & 		$0$ & 	$1.76\cdot 2^{11}$ & 
$1.68\cdot 2^{14}$ & 176 \\
& 200 & $1.07\cdot 2^{19}$ & 	$1.08\cdot 2^{16}$ & 	$1.13\cdot 2^{15}$ & 		$1.72\cdot 2^{12}$ & 	$1.34\cdot 2^{8}$ & 
$1.29\cdot 2^{17}$ & 400\\
\bottomrule
\end{tabular}

\caption{Quantum circuit costs for the circuits we analyze. ``1QC'' are single-qubit Clifford operations and ``M'' are measurements.}\label{table:cipher_costs}
\end{table}

\subsection{Quantum Lookups}\label{sec:qram}
Constructing the initial database from our offline queries requires a QROM\footnote{Also called ``QRACM'' or ``QRAM''.} circuit. We do not assume special, cheap QROM operations (i.e., the QRAM model), but rather give the cost in terms of a Clifford+T simulation of QROM. 

With no depth restriction, the cheapest (in total operation count) is due to Babbush et al.~\cite{PRX:BGB+2018}. Berry et al.~\cite{QUANTUM:BGMMB2019} give a version that is cheaper in T-operations and smoothly parallelizes, but since we have no need to parallelize and consider the full operation count, we use only the Babbush et al. QROM circuit.

\section{Attack circuits and estimates}\label{sec:estimates}
\paragraph{Offline Simon attack.} To estimate the total cost of the attack, we estimated the cost at each value of $u$ and chose the minimum cost, up to some specified limit on $u$. The value of $u$ determines the size of the quantum look-up, which is computed once. We used \autoref{thm:off-rand} to determine the necessary linear system size $m$ and computed the cost to repeat the cipher $m$ times in parallel, based on the cost of a single cipher computation from Q\#. For \Prince{}, which is an FX construction, each parallel repetition needs a copy of the permutation key. However, the permutation key is only infrequently XORed onto the state. With CNOTs, this has depth 1, and can be pipelined efficiently, so we assume the repetitions share the permutation key. This increases the depth by $m$ CNOTs, which is negligible compared to the overall depth of the cipher.

We then estimated the cost of solving an $m\times n$ linear system, using costs from~\autoref{sec:algebra_costs}. Once we found the optimal $m$, we used Q\# to get an exact cost of solving the linear system. The code for this estimation is available \ifeprint{}at \url{https://github.com/sam-jaques/offline-quantum-period-finding/}\else{}on request\fi.

Our results are in~\autoref{table:query_limited_estimate} and~\autoref{table:unlimited_estimate}. We include results for Shor's algorithm to attack RSA-2048 and an exhaustive quantum key search on AES-128 for comparison.

\paragraph{Exhaustive Key Search.} We also estimated the cost of performing an exhaustive quantum key search on the ciphers, summarized in~\autoref{table:grover_estimate}. The circuits for these are slightly different, as we need to attack the full encryption, rather than just the permutation. \Chaskey{} and \Elephant{} modify the key slightly before using it. \Elephant{} transforms the key from 128 bits to the block size, so it is much more efficient to modify the key as part of the search oracle and search a 128-bit space, rather than search a key space as large as the full block size.

To ensure a unique key, we need 2 blocks for \Chaskey{} and 3 blocks for \Prince. We follow the STO approach of~\cite{SAC:DavPri20}, so that we only need to infrequently check blocks besides the first. This also keeps the qubit requirements low; \Prince{} only needs 257 qubits, half of which are only needed as auxiliary qubits for the multi-controlled NOT.

\setlength{\tabcolsep}{6pt}
\begin{table}
\centering
\begin{tabular}{ccccccccc}
\toprule
\multirow{2}{*}{Target} & \multirow{2}{*}{Bitlength} & {Offline} &	\multicolumn{2}{c}{Operations} & \multicolumn{2}{c}{Depth} & \multirow{2}{*}{Qubits} & \multirow{2}{*}{Source}\\
\cmidrule(lr){4-5}\cmidrule(lr){6-7}
 & & Queries & All & T & All & T& & \\
\midrule
RSA & 2048 		& -- & -- 	&$31$ & $31$ & -- & $12.6$ & \cite{ARXIV:GidEke19}\\
\midrule
\Chaskey-8 & 128 & 48 & 64.9 & 64.4 & 56.0 & 53.9 & 14.5 &\multirow{6}{*}{ours} \\
\Chaskey-12 & 128 & 48 & 65.1 & 64.5 & 56.4 & 54.1 & 14.5 &\\
\cmidrule{1-8}
\Prince & 64 & 48 & 65.0 & 64.5 & 55.2 & 53.8 & 14.0& \\
\cmidrule{1-8}
\multirow{3}{*}{\Elephant} & 160 & 47 & 84.1 & 82.5 & 72.6 & 70.4 & 14.8&\\
 & 176 & 47 & 92.5 & 90.9 & 80.8 & 78.5 & 15.1&\\
 & 200 & 69 & 93.6  & 91.7 & 83.7 & 79.3 & 16.4&\\
 \midrule
AES & 128 & 1 & $82.3$ & 80.4 & $74.7$ & 71.6 & $10.7$ &\cite{SAC:DavPri20} \\
\bottomrule
\end{tabular}
\caption{Offline Simon attack cost estimates with the recommended query limits, with RSA and AES for comparison. All figures in log base 2 except bitlength.}\label{table:query_limited_estimate}
\end{table}

\begin{table}
\centering
\begin{tabular}{ccccccccc}
\toprule
\multirow{2}{*}{Target} & \multirow{2}{*}{Bitlength} & {Offline} &	\multicolumn{2}{c}{Operations} & \multicolumn{2}{c}{Depth} & \multirow{2}{*}{Qubits} & \multirow{2}{*}{Source}\\
\cmidrule(lr){4-5}\cmidrule(lr){6-7}
 & & Queries & All & T & All & T& & \\
\midrule
\Chaskey-8 & 128 & 50 & 64.3 & 64.0 & 55.5 & 54.4 & 14.5& \multirow{6}{*}{ours} \\
\Chaskey-12 & 128 & 51 & 64.5 & 64.2 & 55.9 & 55.2 & 14.5 &\\
\cmidrule{1-8}
\Prince & 64 & 50 & 64.4& 64.0 & 55.0 & 54.4 & 14.0 &\\
\cmidrule{1-8}
\multirow{3}{*}{\Elephant} & 160 & 63 & 76.9 & 76.3 & 67.3 & 67.1 & 14.8 &\\
 & 176 & 68 & 82.6  & 81.7 & 72.4 & 72.1 & 15.1 &\\
 & 200 & 76 & 90.7 & 89.7 & 81.1 & 80.1 & 16.4 & \\
 \bottomrule
\end{tabular}
\caption{Offline Simon attack costs without a query limit. All figures in log base 2 except bitlength.}\label{table:unlimited_estimate}
\end{table}

\begin{table}
\centering
\begin{tabular}{ccccccccc}
\toprule
\multirow{2}{*}{Target} & \multirow{2}{*}{Bitlength} & {Offline} &	\multicolumn{2}{c}{Operations} & \multicolumn{2}{c}{Depth} & \multirow{2}{*}{Qubits} & \multirow{2}{*}{Source}\\
\cmidrule(lr){4-5}\cmidrule(lr){6-7}
 & & Queries & All & T & All & T& & \\
\midrule
\Chaskey-8 & 128 & 1 & 80.3 & 77.5 & 79.0 & 75.4 & 8.6 & \multirow{6}{*}{ours} \\
\Chaskey-12 & 128 & 1 & 80.8 & 78.0 & 79.6 & 75.9 & 8.6 &\\
\cmidrule{1-8}
\Prince & 64 & 1.6 & 80.1 & 78.0 & 75.7 & 73.5 & 8.0 & \\
\cmidrule{1-8}
\multirow{3}{*}{\Elephant} & 160 & 0 & 85.1 & 83.1 & 80.2 & 77.3 & 9.6 &\\
 & 176 & 0 & 85.4 & 83.4 & 80.4 & 77.5 & 9.8&\\
 & 200 & 0 & 85.1 & 81.0 & 83.0 & 74.0 & 10.0&\\
 \bottomrule
\end{tabular}
\caption{Attack costs of quantum exhaustive key search using an STO approach. All figures in log base 2 except bitlength.}\label{table:grover_estimate}
\end{table}

\paragraph{Generic collision attacks.} We can remark that in all cases, the total number of quantum gates for the offline Simon's algorithm is close to $2^{n/2-d/6}$, with $2^d$ classical queries, that is, the \emph{query} cost of the generic offline collision attack. This means the offline Simon's algorithm outperforms the generic attack, since its larger polynomial factor is not an issue for cryptographic parameter sizes.

\section{Conclusion}

\subsubsection{A new kind of attack.} Quantum exhaustive key search may not be a real threat to symmetric cryptography because of its poor parallelization~\cite{PRA:Zalka99,EC:JNRV20} and the expected overheads of error correction. However, we showed that there are other avenues of quantum attack that may be more feasible. For example, \Chaskey{} and \Prince{} have ``only'' ${33}$ more bits of quantum security than RSA-2048, widely believed to be completely broken in a post-quantum setting. 

Comparing the security of RSA-2048 to \Chaskey{} and \Prince{}, we point out that our attack requires less than 4 times as many logical qubits, but many more quantum operations. This means breaking these ciphers will take much longer and require much more coherence than breaking RSA. However, adding more coherence to an already-coherent quantum computer is relatively easy. For surface code error correction, coherence grows exponentially with code distance, and the qubit overhead grows only quadratically~\cite{PRA:FMMC2012}. Moreover, our attacks tend to have a lower depth than quantum search, which may also help its implementation. Thus, we believe that these attacks could be an interesting milestone for quantum computers, much harder than RSA-2048 factoring, but much easier than AES-128 key recovery. 


\subsubsection{On quantum-safe symmetric cryptography.} We found that \Chaskey{} (independently of its number of rounds) and \Prince{} have almost identical quantum security. Moreover, the data limitation of Chaskey has a negligible impact on the attack cost and our attacks end up being almost a million times cheaper than the corresponding quantum key search.

Our attack on \Elephant{} is less competitive and requires more quantum operations than the direct key search. This is mainly because our attack targets the state size, and \Elephant{}'s key size is smaller. The data limitation also slows our attack, but the cost increase is much smaller than the cost increase of the classical attack. Moreover, this attack shows that to make an \Elephant{} instance with significantly more quantum security than $2^{64}$ queries would require an increase in both the key and the state length. One of \Elephant{}'s features compared to other lightweight cryptography candidates is its small state size, so such a change would make it less competitive.

To counteract the offline Simon attack and to achieve quantum security, we recommend:
\begin{itemize}
\item
Using a large \emph{state} size, not just a large key size.
\item
Not relying on data limits, as these have limited impact on quantum attacks.
\item
Avoiding the Even-Mansour and FX constructions altogether.
\end{itemize}

For an example of the last idea, the design of the recent \Prince{} v2~\cite{SAC:BEKLLMNRTW20} is very close to the original \Prince, but with a simple key schedule that replaces the FX construction.


\subsubsection{Immediate implications.} We stress that, like quantum exhaustive key search or factoring, a patient attacker could apply this attack to today's communications, as it is an offline attack: the data can be collected before any quantum computation.

This is especially important for lightweight cryptography, which is intended for use in embedded systems, RFID chips or sensor networks, where an update is either impractical or downright impossible.

\ifeprint
\subsubsection{Acknowledgements.} The authors would like to thank Léo Perrin for fruitful discussions about S-boxes. Samuel Jaques was supported by the University of Oxford Clarendon fund.
\fi
\bibliographystyle{splncs03}
\bibliography{cryptobib/abbrev3,cryptobib/crypto,biblio}
\end{document}